\RequirePackage{fix-cm}
\documentclass[letterpaper,10pt,journal]{IEEEtran}
\IEEEoverridecommandlockouts

\usepackage{cite}
\usepackage{amsmath,amssymb}
\usepackage{booktabs}
\usepackage{multirow}
\usepackage{graphicx}
\usepackage{cuted}
\usepackage{capt-of}
\usepackage{xcolor}
\usepackage{url}
\usepackage[hidelinks]{hyperref}
\usepackage{xspace}
\makeatletter
\providecommand{\@setmarks}{}
\makeatother

\newcommand{\cpt}{\textsc{CPT}\xspace}

\newcommand{\best}[1]{\textbf{#1}}

\newif\ifralappendices
\ralappendicestrue
\newif\ifinlinetables
\inlinetablesfalse

\title{OmniRAS: Standardizing Foundation Model Training and Evaluation in
Robot-Assisted Surgery}
\author{%
Leonardo Borgioli$^{\dagger,*}$,
Neil Getty$^{\ddagger,*}$,
Wenli Xiu$^{\P}$,
Jessica Cassiani$^{\S}$,
Alvaro Ducas$^{\S}$,
Carlos Agustin Orda$^{\S}$,
Hira Waris$^{\S}$,
Fangfang Xia$^{\ddagger}$,
Rick Stevens$^{\ddagger}$,
Pier Cristoforo Giulianotti$^{\S}$,
and Milos Zefran$^{\dagger}$
\\[0.5em]
{\small
$^{\dagger}$Department of Electrical and Computer Engineering,
University of Illinois Chicago, Chicago, IL, USA
}\\[-0.1em]
{\small
$^{\ddagger}$Argonne National Laboratory, Lemont, IL, USA
}\\[-0.1em]
{\small
$^{\S}$Department of Surgery,
University of Illinois Chicago, Chicago, IL, USA
}\\[-0.1em]
{\small
$^{\P}$Affiliated Hospital of Qingdao University, Qingdao, China
}\\[0.2em]
{\small
$^{*}$These authors contributed equally.
}
}

\hypersetup{
  pdfauthor={Leonardo Borgioli, Neil Getty, Wenli Xiu, Jessica Cassiani,
  Alvaro Ducas, Carlos Agustin Orda, Hira Waris, Fangfang Xia,
  Rick Stevens, Pier Cristoforo Giulianotti, Milos Zefran}
}
\begin{document}
\maketitle

\begin{abstract}

Few foundation models exist for robot-assisted surgery, partly because large robotic-surgery video corpora are difficult to assemble and existing models are evaluated mostly on laparoscopic benchmarks.  Further, most existing models are evaluated on a small set of public benchmarks, mostly focused on laparoscopic surgery. We present OmniRAS, a family of 1B- and 2B-parameter V-JEPA-2.1 encoders for robot-assisted surgery, and detail their training. First, we release two densely annotated robotic-cholecystectomy datasets: OmniRAS-PR and a multi-label YT-Chole tool--verb--target task, the first triplet-style annotation for robotic cholecystectomy, together with splits, probe protocols, and an inter-rater study validating the shared phase ontology. Second, we document continued pretraining at up to 256 compute nodes with global batch 6,144 over 19 sources totaling approximately 2,650 hours of surgical video, 51\% robotic, and analyze compute and data composition. Third, we evaluate against raw V-JEPA-2.1 and specialized surgical models on six tasks spanning triplet, phase, and step recognition, action segmentation, and detection, under frozen-encoder and final-four-block fine-tuning regimes. Across three seeds, this yields 254 downstream runs, including 109 with partial backbone fine-tuning. The best OmniRAS models achieve the strongest adapted results across all task families, while frozen differences are smaller.

\end{abstract}


\begin{strip}
\centering
\vspace{-14pt}
\includegraphics[width=\textwidth,trim=8mm 8mm 7mm 9mm,clip]{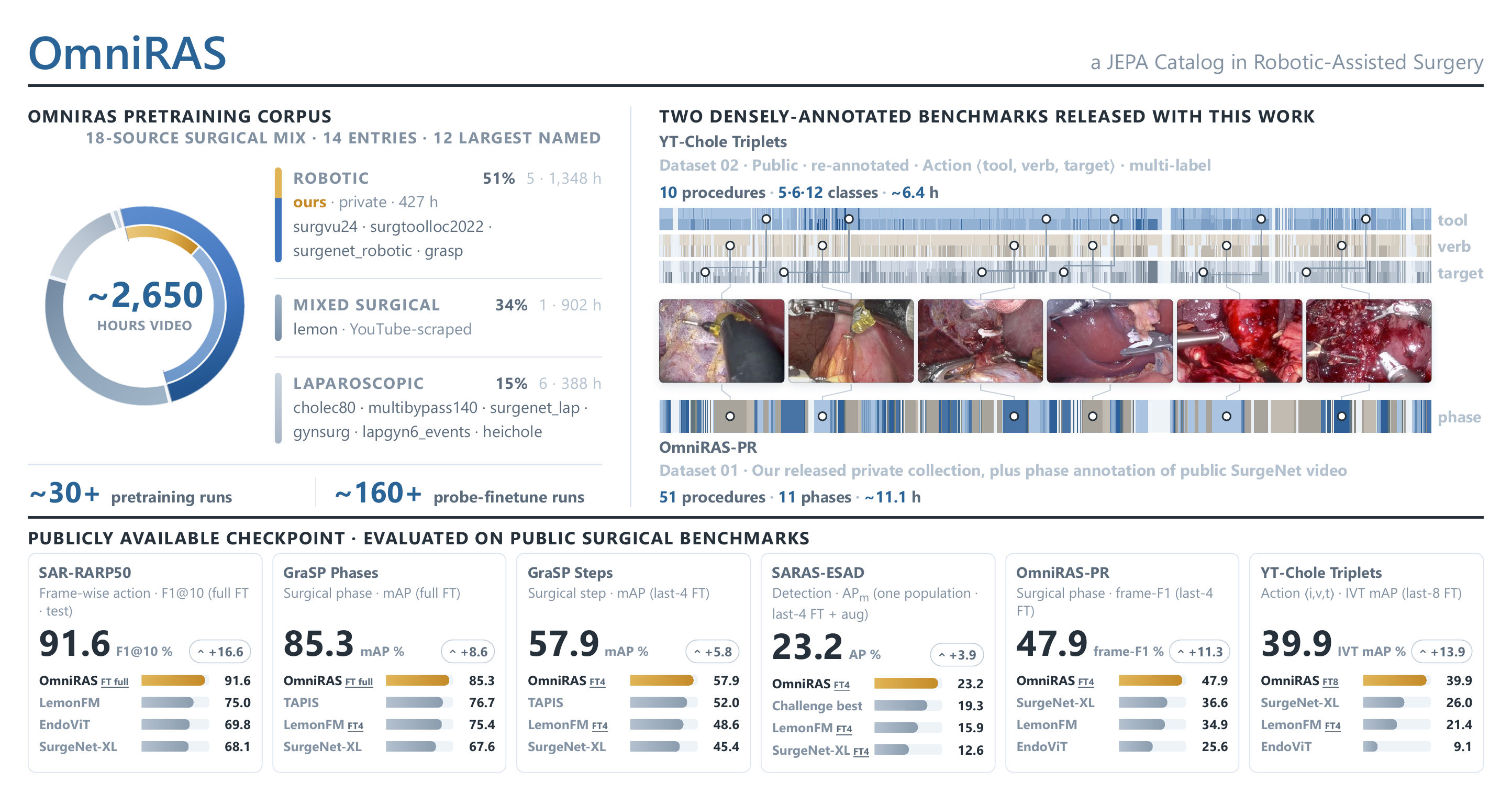}
\captionof{figure}{\textbf{Overview of OmniRAS. Billion-scale V-JEPA-2.1 encoders} are
continued-pretrained at up to 256 nodes on a 19-source, approximately
2,650-hour surgical-video catalog that is 51\% da Vinci robotic video by hours, with rehearsal sources varying by production run, and evaluated on the two densely
annotated robotic-cholecystectomy datasets released with this work together
with the public SAR-RARP50, GraSP, and SARAS-ESAD benchmarks. The bottom row reports representative results, each corresponding
to an entry in the appendix tables; complete protocols, matched controls, and
statistical qualifications are given in the main text and appendices.}
\label{fig:modelcard}
\vspace{-5pt}
\end{strip}

\section{Introduction}
Robot-assisted surgery (RAS) produces long, information-dense videos in which clinically relevant events unfold at very different spatial and temporal scales. A useful representation must preserve instruments and anatomy, capture short surgical actions, and retain enough temporal context to describe procedural phases. Surgical video is, therefore, a natural setting for large predictive video models, where abundant unlabeled footage can be exploited before expensive task-specific annotations are introduced.

Recent self-supervised methods learn transferable visual features through pixel reconstruction or prediction in representation space~\cite{he2022mae,assran2023ijepa,tong2022videomae}. DINOv3 provides strong global and dense image representations~\cite{oquab2023dinov2}, while V-JEPA~2 scales predictive learning to video~\cite{assran2025vjepa2} and V-JEPA-2.1 introduces dense objectives and deep self-supervision~\cite{mur2026vjepa21}. Surgical pretraining has followed the same direction, from procedure-specific SurgMAE~\cite{jamal2023surgmae} to broader models including GSViT, SurgeNetXL, SurgVISTA, and LemonFM~\cite{schmidgall2024gsvit,jaspers2025surgenet,yang2026surgvista,che2026lemon}. These works show that large surgical encoders can transfer across tasks, but they do not yet provide a common way to determine what properties of the representation are actually improved. This matters because the quality of surgical representation is not captured by a single benchmark. Workflow recognition probes temporal structure over seconds to minutes and has evolved from frame- and tool-based models such as EndoNet to explicit temporal refinement in TeCNO, TMRNet, Trans-SVNet, and LoViT~\cite{twinanda2017endonet,czempiel2020tecno,jin2021tmrnet,gao2021transsvnet,liu2023lovit}. Fine-grained activity recognition instead focuses on shorter interactions, commonly through the instrument--verb--target factorization standardized by CholecT50 and the CholecTriplet challenge~\cite{nwoye2022splits,nwoye2022challenge}. Detection and other spatially resolved tasks test whether localized information survives pretraining. Performance on one of these axes does not imply performance on the others, so evaluating a foundation model requires more than a single downstream score.

For RAS, the problem is more pronounced. Even frontier general-purpose
multimodal models remain far from competitive on fine-grained surgical understanding: in our few-shot YT-Chole evaluation, a frontier multimodal model (Claude Opus 5)
achieves only 0.139 IVT mAP, compared with 0.426 for our task-adapted
representation (Appendix~\ref{tab:opus5-ytchole}). Most benchmarks used to evaluate surgical foundation models are laparoscopic, leaving robotic encoders
to be judged largely off-domain. Comparisons also frequently change the
downstream architecture together with the backbone, making it difficult to
determine whether a gain comes from the learned representation or from the
readout used to expose it. Finally, production-scale surgical pretraining requires choices about dataset composition, sampling, masking, and compute allocation, yet the effects of these decisions are rarely studied systematically and reported.

OmniRAS addresses these gaps by establishing a common evaluation framework centered on robot-assisted surgery. We introduce two annotated robotic-cholecystectomy benchmarks: YT-Chole Triplet, derived from SurgeNet videos~\cite{jaspers2025surgenet}, and OmniRAS-PR, a phase-recognition benchmark combining private and SurgeNet procedures under a unified 11-class ontology. Together, they enable evaluation in the fully robotic (as opposed to laparoscopic or open surgery) domain at both fine-grained activity and procedural-workflow scales. We complement them with existing robotic benchmarks, including GraSP~\cite{ayobi2024grasp}, SAR-RARP50~\cite{psychogyios2024sarrarp50}, and SAR-ESAD~\cite{bawa2021esad} to test representation transfer across temporal and spatial tasks. Within this framework, we study which factors in surgical \cpt affect transfer and whether the resulting encoders provide stronger shared perception backbones for robot-assisted surgery than raw and specialized surgical baselines under matched readouts.

Our contributions are:
\begin{itemize}
    \item \textbf{Two annotated benchmarks.} We define and release two complementary custom evaluations of robotic cholecystectomy: OmniRAS-PR, a 51-procedure phase-recognition benchmark combining 41 private and 10
SurgeNet procedures under a shared 11-class ontology, and YT-Chole action
triplets with independent tool, verb, and target axes, the first multi-label
triplet task on robotic cholecystectomy video.
    \item \textbf{A validated phase and triplet ontology.} Two independent re-annotations against the existing reference yield substantial agreement across all three rater pairs. 
    \item \textbf{A documented production-scale pretraining campaign.} We report continued pretraining at up to 256 nodes with a global batch of 6,144, including three production runs of 9.22 M, 18.43 M, and 36.86 M samples and exploratory screening over budget, catalog composition, sampling, and masking objective. The longest production configuration yields the strongest transfer, although budget and catalog are not fully disentangled; we report the campaign as practical guidance rather than a scaling law.

    \item \textbf{A common probe of every encoder.} We evaluate 1\,B and 2B
    V-JEPA-2.1 encoders, raw and our surgical (OmniRAS), alongside SurgeNetXL, LemonFM, and
    EndoViT on YT-Chole triplets, OmniRAS-PR phases, GraSP phase and step, SARAS-ESAD detection, and SAR-RARP50 action recognition, under frozen probing and
    last-four-block fine-tuning. To our knowledge, this is the first such
    comparison conducted entirely on robot-assisted surgery across 3-seeds, with no
    laparoscopic benchmark standing in for the target domain. The adapted surgical
    encoders achieve the strongest results across the task families evaluated.
    \item \textbf{A label-free test of objective-aligned predictive fidelity.} We replicate the
    V-JEPA-2.1 masking objective as an evaluation, with no task head, and show
    that the surgical checkpoint predicts masked targets in its own latent
    space more accurately than its initialization on 60 of 60 paired unit-level comparisons across five evaluation settings, a paired, chance-corrected
    measurement of continued pretraining independent of any probe.
    \item \textbf{A complete record of controls.} We document
    checkpoint-selection, seed, launcher, and aggregation controls, including
    negative results and one protocol defect that invalidated an earlier set of
    conclusions. Across three production continued-pretraining runs totaling 3,038 node-hours and 254 downstream runs, including 109 with partial backbone fine-tuning, we quantify downstream variability across seeds.
\end{itemize}

\section{Datasets and Evaluation Tasks}

\begin{table*}[!h]
\centering
\caption{Dataset and probe overview. Counts are clips for the custom tasks and
approximately 1-second temporal tokens for SAR-RARP50 Action. }
\label{tab:datasets}
\renewcommand{\arraystretch}{1.12}
\setlength{\tabcolsep}{4.2pt}
\resizebox{\textwidth}{!}{%
\begin{tabular}{lllcrrrl}
\toprule
\textbf{Task} & \textbf{source} & \textbf{prediction head} & \textbf{classes} &
\textbf{train} & \textbf{val} & \textbf{test} & \textbf{primary metric} \\
\midrule
\multirow{2}{*}{OmniRAS-PR}
& Our-private & learned-query attentive probe & 11 & 12,815 & 3,368 & -- & macro-F$_1$ \\
& SurgeNet~\cite{jaspers2025surgenet} & learned-query attentive probe & 11 & 2,704 & 679 & -- & macro-F$_1$ \\
YT-Chole Triplets & SurgeNet~\cite{jaspers2025surgenet} & token-pooled multi-task sigmoid heads (tool/verb/target) & 5/6/12 & 3,537 & 1,696 & -- & IVT mAP \\
SAR-RARP50 Action~\cite{psychogyios2024sarrarp50} & public & global self-attention $+$ residual dilated temporal convolutions, best of three LRs & 8 & 272,472 & 40,560 & 78,048 & F$_1$@10 \\
GraSP Phases~\cite{ayobi2024grasp} & public & SAR-RARP50 head transferred unchanged (class count and class weights only), best of three LRs & 11 & 9,191 & train fold & 5,354 & phase mAP \\
GraSP Steps~\cite{ayobi2024grasp} & public & SAR-RARP50 head transferred unchanged (class count and class weights only), best of three LRs & 21 & 6,031 & train fold & 3,555 & step mAP \\
SARAS-ESAD Detection~\cite{bawa2021esad} & public & temporal presence classifier $+$ class-conditioned box regressor & 21 & 2,468 & 380 & 147 & detection mAP \\
\bottomrule
\end{tabular}}
\end{table*}

\ifralappendices
\begin{table*}[tb]
\centering
\caption{Label inventory for the three custom corpora. The two phase datasets
share one aligned eleven-class ontology and are listed together; YT-Chole
Triplets uses independent multi-label axes.}
\label{tab:labels-main}
\renewcommand{\arraystretch}{1.12}
\setlength{\tabcolsep}{5pt}
\begin{tabular}{lllp{10.1cm}}
\toprule
\textbf{corpus} & \textbf{label type} & \textbf{number} & \textbf{classes} \\
\midrule
OmniRAS-PR & single-label phase & 11 & exposure of working area; retraction of gallbladder neck; opening anterior peritoneal layer (Calot); opening posterior peritoneal layer (Calot); isolation of cystic duct; isolation of cystic artery; clipping of cystic duct; clipping of cystic artery; division of cystic duct and artery; dissection of gallbladder from liver; specimen retrieval \\
\midrule
\multirow{3}{*}{YT-Chole Triplets} & \multirow{3}{*}{multi-label triplet} & 5 tools & clipper; grasper; hook; irrigator; scissors \\
& & 6 verbs & coagulation; grasp/retract; cut/dissect; clean; clip; null \\
& & 12 targets & connective tissue; cystic duct; adhesion; cystic pedicle; gallbladder; liver; peritoneum; cystic artery; falciform ligament; omentum; fluid; null \\
\bottomrule
\end{tabular}
\end{table*}
\fi

\subsection{Two annotated custom benchmarks}
\label{sec:custom}

The first contribution of this work is a set of two densely annotated
robotic-cholecystectomy evaluations, released together with their split
definitions and probe protocols. They exist because the public surgical corpora
available to us cover a narrow band of procedures and, for cholecystectomy in
particular, offer no multi-label interaction annotation on robotic video. The released annotations span complementary temporal scales: phase labels capture procedural state under a shared coarse ontology, while YT-Chole Triplets captures short interactions at a finer granularity.

\textbf{OmniRAS-PR} is a robotic cholecystectomy phase-recognition dataset comprising 51 procedures and approximately 11.1 hours of densely annotated video. It combines 41 private procedures (approximately 7.3 hours) with ten publicly available procedures from SurgeNet~\cite{jaspers2025surgenet} (approximately 3.8 hours), all annotated under the same 11-class procedural-phase ontology. The private portion contains 156,217 training and 41,108 validation clips, while the public portion holds out two complete procedures for evaluation, preventing clip-level leakage. This unified construction provides directly comparable phase annotations across private and public robotic cholecystectomy data.

\textbf{YT-Chole Triplets} re-annotates the same ten batches averaging 3 procedures,
approximately 6.4 hours, as a multi-label
$\langle$tool, verb, target$\rangle$ prediction problem with five tool classes,
six verbs, and twelve targets. Each axis carries an independent sigmoid head
because several labels can be active in one clip, and the joint IVT metric is
macro average precision over supported triplets rather than over the full
Cartesian product. To our knowledge this is the first triplet-style action
annotation defined on robotic cholecystectomy video.

Table~\ref{tab:datasets} summarizes the custom and public task sizes together
with  corresponding probe definitions\ifralappendices, and
Table~\ref{tab:labels-main} lists the full label inventory\fi. YT-Chole Triplet and
 OmniRAS-PR phase probes results are reported in Section~\ref{sec:results}.
Section~\ref{sec:irr} reports an inter-rater study showing that the shared
eleven-class phase ontology is reproducible by independent experts.

\subsection{Annotation Protocol}
The annotation protocol was developed using the cholecystectomy chapter of \cite{gruessner2024cholecystectomy} as the main procedural reference. The chapter provides a structured description of the surgical procedure, including its main steps, actions, and relevant anatomical structures, which were used as anchors for defining the annotation triplets. Before starting the annotation process, an initial consensus phase was conducted to agree on the annotation ontology and clarify the definitions of potentially ambiguous surgical actions, such as distinguishing \textit{coagulation} from \textit{dissection}. Initial annotations were then performed by three surgeons. Two additional re-annotators subsequently reviewed the annotations to verify their correctness, ensure consistency with the agreed ontology, and resolve any remaining disagreements.

\subsection{Annotation reliability of the phase ontology}

The eleven-class ontology is shared by both phase corpora, so the extent to which independent experts reproduce it determines the utility of the corpus.
A 10\% sample of YT-Chole, 39.5 minutes at 4\,fps (9,480 frames), was
independently re-annotated for phase by two raters (A1--A2) against the existing reference annotation, all using the same protocol
and label vocabulary. Annotation is sparse by design: raters label the
intervals they can identify and leave ambiguous video unlabeled, so agreement
is computed frame-wise over the frames labeled by both members of a pair, since
scoring unlabeled frames as a background class would inflate it.

\begin{figure*}[tb]
\centering
\includegraphics[width=\linewidth]{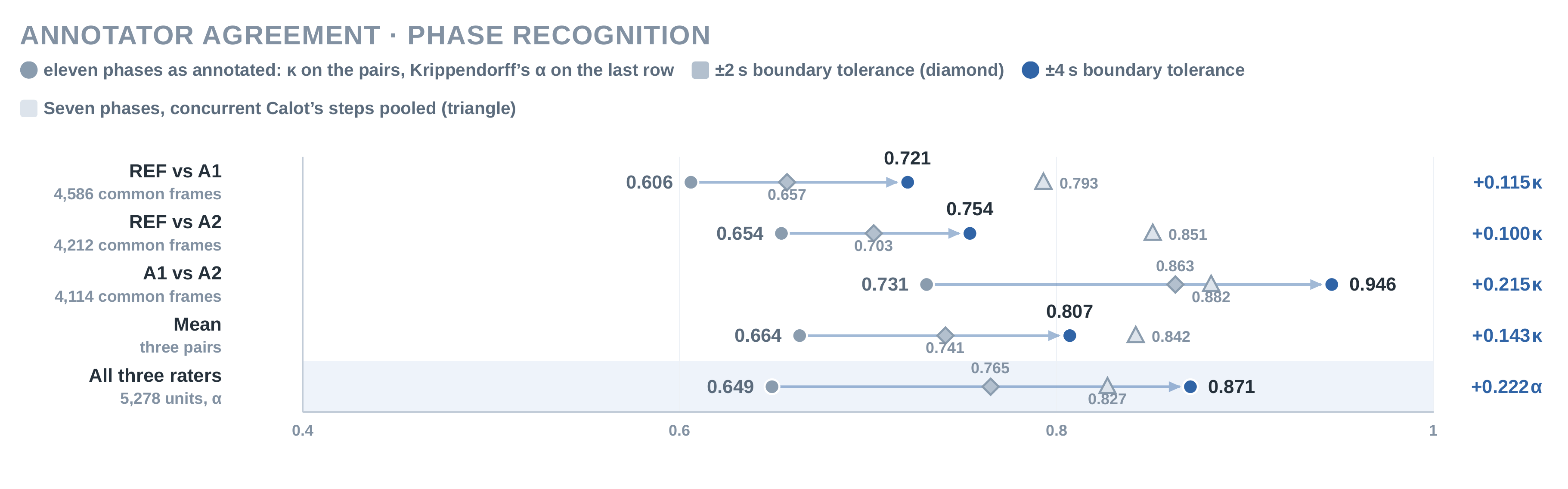}
\caption{\textbf{Annotation quality of the eleven-class phase ontology.} Measured on a
10\% sample of YT-Chole re-annotated by two raters against the existing
reference.
Every pair independently reaches substantial agreement at eleven classes. The
arrow traces the boundary-tolerance ladder: zero tolerance (circle), $\pm2$\,s. 
(diamond) and $\pm4$\,s (head). At four seconds of tolerance the mean reaches
$\kappa=0.807$ over the full eleven classes, with no distinction removed from
the label set; this is the operating point that characterizes the ontology. The
open triangle is the seven-class $\kappa$ obtained by pooling the five
concurrent Calot's steps, at zero tolerance. It lies beyond the arrowhead in
every pair except A1--A2, indicating that the residual disagreement arises from
raters applying two valid descriptions to one concurrent activity rather than
from mistimed boundaries.}
\label{fig:irr}
\end{figure*}

We score at three levels of boundary tolerance, discarding frames within
$\pm\epsilon$ of any segment boundary in \emph{either} annotation before
rescoring the remainder (Figure~\ref{fig:irr}). At zero tolerance, the mean pairwise Cohen's $\kappa$ over all eleven classes is 0.664, \emph{substantial} on the Landis--Koch scale
and reached independently by every one of the three pairs; it rises to 0.741 at
$\pm2$\,s and 0.807 at $\pm4$\,s. The gain is not an artifact of the smaller
frame count, since dropping as many frames at random (50 seeded draws per pair)
recovers the unmasked $\kappa$.

The residual disagreement is structural rather than diffuse: 82.1\% of it
involves the five-phase Calot's-triangle cluster, whose sub-activities are
performed concurrently rather than in sequence, while phases outside it are
annotated consistently. Pooling the cluster raises the mean to 0.833, and that
coarser view is released alongside the full ontology; all model results use the eleven-class protocol and should be interpreted in light of the observed inter-rater agreement (k=0.807 at $\pm4$s), rather than perfect agreement. Model-to-model comparisons are unaffected,
since every encoder is scored against the same reference. 

\subsection{Annotation reliability of the triplet ontology}
\begin{figure*}[!h]
\centering
\includegraphics[width=\linewidth]{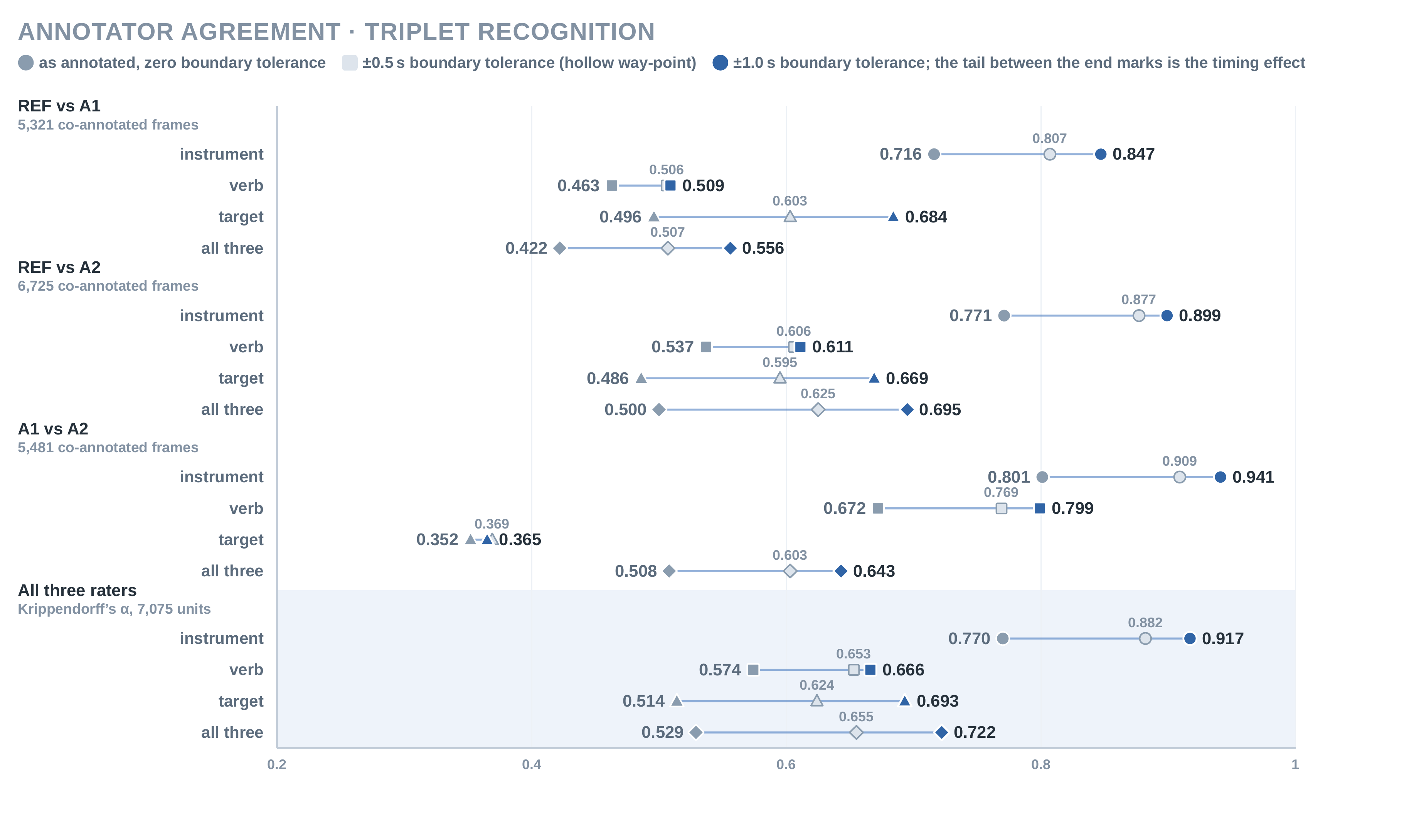}
\caption{\textbf{Annotation quality of the action-triplet ontology.} measured on the
frames of YT-Chole co-annotated by two raters against the existing reference.
Each lane is one label axis --- instrument (circle), verb (square), target
(triangle), and all three simultaneously (diamond) --- with the grey mark at
zero boundary tolerance and the blue mark at $\pm1.0$\,s; the tail between them
is the timing effect. Instrument is substantial to almost perfect on every pair
and is not limited by timing, whereas target is the axis that limits the
protocol. Its A1--A2 value is depressed by one category covering 76\% of frames
rather than by disorder: Gwet's AC1 reads 0.755 on the same frames. The shaded
band reports Krippendorff's $\alpha$ over all three raters at once, which needs
no averaging over pairs.}
\label{fig:irr-triplet}
\end{figure*}

The triplet ontology is independent of the phase ontology and is annotated along three separate axes, so its reliability was assessed separately. The same 10\% sample was re-annotated for action triplets (instrument,$|$,verb,$|$,target) by the same two raters against the same reference, using the same sparse-labeling protocol. Because triplets are multi-label, each frame was reduced to its longest-active label on each axis before scoring. Agreement was evaluated at zero, $\pm0.5$,s, and $\pm1$,s boundary tolerance (Figure~\ref{fig:irr-triplet}). At zero tolerance, mean pairwise Cohen's $\kappa$ was 0.763 for instrument, 0.557 for verb, 0.445 for target, and 0.477 for the complete triplet. These values improve substantially once a small temporal tolerance is allowed. At $\pm0.5$,s, the two-rater Krippendorff's $\alpha$ already reaches 0.882 for instrument, 0.653 for verb, 0.624 for target, and 0.655 for the full triplet, increasing further to 0.917, 0.666, 0.693, and 0.722 at $\pm1$,s. The half-second tolerance is especially informative because it captures most of the improvement for instrument and verb, showing that much of the apparent frame-level disagreement comes from slightly different boundary placement rather than different interpretations of the action. The random frame-removal control supports this interpretation, since removing the same number of frames at random does not reproduce the increase in agreement.

Overall, the results support temporal consistency of the triplet ontology under the dominant-label projection used for scoring. The remaining differences are small and mostly temporal, rather than evidence of a systematic disagreement in the ontology. An a-priori grouping of related verbs and anatomical targets produces only modest additional gains compared with random partitions of the same size, so there is little evidence that collapsing the vocabulary would provide a more reliable representation. The main exception is the A1--A2 target comparison, where $\kappa=0.352$ despite 76.4\% raw agreement; however, one category accounts for 75.9\% of frames, making Cohen's $\kappa$ particularly sensitive to prevalence imbalance, while Gwet's AC1 reaches 0.755. This isolated case, therefore, does not change the broader pattern. In practical terms, agreement is already strong with a tolerance of only $\pm0.5$,s, suggesting that the dataset captures the triplet ontology consistently and that the small residual variation mainly reflects the difficulty of placing action boundaries at exactly the same frame.

We plan to release both datasets, reannotations, and the surgeons' background associated with each annotation when possible. 

\label{sec:irr}

\subsection{Public benchmarks}

The public benchmarks were selected under two constraints. The first is the domain.
A claim about robot-assisted surgery cannot be supported by laparoscopic
evaluation, since the two settings differ in camera motion, instrument
appearance, and procedural workflow; we therefore admit only benchmarks
recorded on robotic platforms and exclude the laparoscopic corpora on which surgical foundation models are usually scored. The second is task coverage. A single benchmark family probes only one property of the representation, and performance on one does not imply performance on another. Our evaluation, therefore, spans three complementary axes of surgical video understanding: coarse temporal structure over minutes, fine temporal structure over seconds, and sparse spatial structure at the object level. We operationalize these axes with three benchmarks covering three task types across two procedures.

\textbf{SAR-RARP50}~\cite{psychogyios2024sarrarp50} comprises fifty robot-assisted
radical prostatectomy videos of the suturing phase, released with temporal
action labels. Action recognition uses eight gesture classes
and requires temporally coherent predictions over the sequence; we report the
standard segmental F$_1$@10 on the official test videos.

\textbf{GraSP}~\cite{ayobi2024grasp} annotates radical prostatectomy at two levels of a
single workflow hierarchy: eleven long phases spanning minutes, and twenty-one
short steps within them. Evaluating both on the same videos and the same
official partition isolates temporal granularity from every other factor, since
only the label horizon changes between the two. We report mAP on the official
test partition for both levels, The learning rate is selected on a fold held out from the official training cases.

\textbf{SARAS-ESAD}~\cite{bawa2021esad} evaluates surgeon action detection on
prostatectomy video with twenty-one classes, and is the only benchmark
considered here that requires localization rather than recognition alone.
Predictions are class-conditioned bounding boxes. We report challenge AP
together with AP10, and AP50; presence mAP, mean IoU, and box mAP@50 are
retained as diagnostics that separate recognition from localization.

The custom corpora of Section~\ref{sec:custom} complete the coverage on the
procedure axis. The public benchmarks are prostatectomy, whereas YT-Chole and
OmniRAS-PR are cholecystectomy, so a result that holds on both is not a property of a
single procedure.

\subsection{Pretraining data}

The 2B checkpoint is a ViT-giant trained with the V-JEPA-2.1 self-supervised
objective, initialized from the publicly released general-video weights and
continued on a domain-focused surgical corpus. Pretraining clips are sixteen
frames sampled at four frames per second, a temporal window of approximately
four seconds, cropped at high resolution, and streamed through a WebDataset
pipeline. The corpus is a multi-source collection aggregated into a single sharded pool. The majority is surgical and spans both robotic and
laparoscopic domains across several procedure types. The robotic sources are
our private contribution (427h), SurgToolLoc/SurgVU, GraSP (robotic
radical prostatectomy), and a scrubbed SurgeNet robotic subset. The
laparoscopic and mixed-modality sources are Cholec80, HeiChole, MultiBypass140
(gastric bypass), the gynecologic sets GynSurg and LapGyn6, a laparoscopic
SurgeNet subset, the multi-procedure LEMON/Surg-3M collection, and a bundle of
smaller established benchmarks (JIGSAWS, SurgVisDom, CRCD, and several
EndoVis/MICCAI instrument sets). To this surgical core, the catalog adds two
non-surgical anchors: Kinetics-400, which provides a general-video rehearsal
signal against the forgetting of general-action representations that we measure
under surgical-only continued pretraining, and the Open-H-Embodiment
medical-robotics dataset.

Source proportions are not hand-tuned. Every source enters with equal nominal
weight and the mixture is balanced at draw time by a square-root
inverse-frequency temperature sampler, which prevents high-volume sources such
as Kinetics-400 from dominating the batch and keeps the effective distribution
predominantly surgical with a bounded general-video fraction. Quality filters
discard near-static and black clips.

\section{Models and Protocols}

\subsection{Encoders and continued pretraining}

V-JEPA is a self-supervised video model that learns by prediction in
representation space rather than in pixel space. A clip is split into a visible
context and a hidden target, and an online encoder with a small predictor must
reproduce the target encoder's latent description of the hidden region from the
context alone. Predicting representations rather than pixels frees the model
from reconstructing appearance it cannot infer, such as specular highlights or
smoke, so capacity goes to what is predictable: in surgical video, largely how
instruments and anatomy are arranged and how that evolves. No labels enter at
any point. V-JEPA-2.1 adds a dense per-token objective and deep self-supervision
over several encoder depths to the V-JEPA~2
recipe~\cite{assran2025vjepa2,mur2026vjepa21}; we continue that identical
objective on surgical video, which is what \cpt{} denotes throughout.

A consequence of this formulation is that the pretraining $\ell_1$ loss is not,
by itself, a meaningful measure of representation quality: it can decrease as
the learned latent space evolves without implying better downstream features.
Probes are therefore essential, providing an external, task-grounded measure of
whether continued self-supervision is actually improving the representation.

We evaluate the released V-JEPA-2.1 ViT-g and ViT-G encoders, with widths
1,408 and 1,664 and depths 40 and 48, respectively. Including the predictor,
these configurations contain approximately 1B and 2B parameters. Raw'' denotes the released general-domain checkpoint; \cpt'' denotes continued
self-supervised pretraining on a multi-source surgical-video catalog. The
headline 1B model is epoch~19.

The central comparison is not a scaling study. It quantifies the effect of
surgical exposure on transfer relative to the matched raw initialization.

\subsection{Probes}
\begin{figure}
    \centering
    \includegraphics[width=1\linewidth]{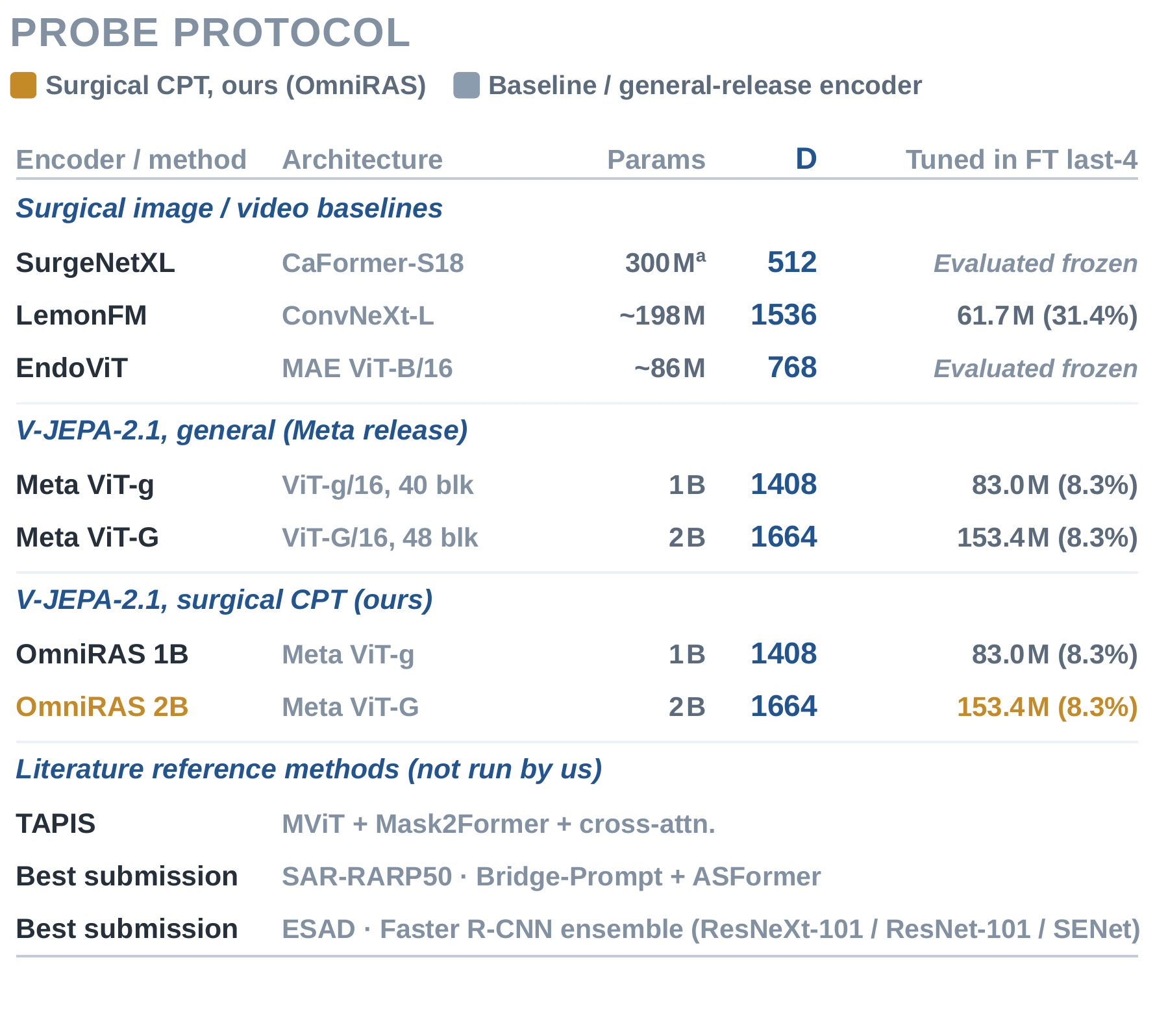}
    \caption{Encoder roster and probe adaptation protocol. $D$ denotes the token width inherited by the task head; FT4 unfreezes the final four encoder blocks and is depth-matched rather than parameter-matched. Literature methods are
reported for reference only. Surgical video foundation models GSViT, SurgVISTA, SurgMAE are not benchmarked: their pretraining corpora are laparoscopic rather than robotic. The surgical baselines evaluated here are therefore image encoders, and comparisons between them and the video encoders should be read accordingly.}
    \label{fig:probe-protocol}
\end{figure}

A pretrained encoder emits tokens, not phases or triplets, so it cannot be evaluated directly: a trainable readout must map its representation onto the target label space, and that choice affects what the resulting score measures. Fine-tuning the encoder end-to-end allows task adaptation to reshape much of the representation, testing the adaptation budget as well as the quality of pretraining. A frozen encoder with a lightweight trained head provides the complementary view: whatever the head recovers must already be accessible from a representation it cannot alter. This setting also corresponds to the shared-backbone deployment regime of interest, in which one cached encoder can serve multiple downstream tasks. We therefore use frozen evaluation as the common representation-level comparison and, where supported, pair it with a controlled partial fine-tuning regime (FT4). In the frozen setting, the pretrained encoder is held fixed, and only the task head is optimized on its spatio-temporal tokens. In FT4, the same head, objective, data split, and optimization protocol are retained while only the final four encoder blocks are unfrozen, using an encoder learning rate of $10^{-5}$. This limited adaptation tests whether task-relevant information that is not readily accessible to a shallow readout can be exposed without allowing unrestricted end-to-end specialization. Figure~\ref{fig:probe-protocol} summarizes the encoder roster and the adaptation budget associated with this protocol. FT4 is deliberately \emph{depth-matched}, not parameter-matched: four blocks are unfrozen for each compatible encoder, so the number and fraction of trainable backbone parameters vary with architecture. Likewise, the probe head inherits the encoder token width $D$; the head design and objective are matched across encoders, but its absolute parameter count therefore scales with the representation dimension. For the Lemon model, we decided to apply this advantage in the evaluation to compensate for the difference in parameters.  Encoders for which only frozen evaluation is performed are marked accordingly in Fig.~\ref{fig:probe-protocol}, while literature reference methods are shown for context but are not rerun under our protocol. The task heads fall into three structural families: a clip-level attentive or pooled classifier, a temporal action-segmentation head, and a spatially resolved detection head. These are instantiated across the five surgical benchmarks described below. Within each probe, frozen and FT4 comparisons use otherwise matched task definitions and training protocols so that the effect of backbone adaptation can be isolated from changes to the downstream readout.

\textbf{Surgical action-triplet recognition (YT-Chole).} The action triplet is
factorized into three independent multi-label sub-problems: instrument, verb,
and target. The encoder tokens are reduced to a single clip-level descriptor,
by pooling either the token features or the per-token predictions, and passed
to a shared multi-task head that emits one multi-label branch per sub-problem,
each trained with binary cross-entropy. We report class-averaged mAP separately
for instrument, verb, and target, together with their arithmetic mean. IVT mAP
measures the harder joint recognition problem: triplet confidence is obtained
from the product of the three marginal probabilities under an independence
assumption, and AP is then computed over the supported
instrument--verb--target combinations rather than the full Cartesian product.

\textbf{Phase recognition (OmniRAS-PR).} Phase recognition uses an attentive
classifier. A single learnable query token cross-attends over the full set of
encoder tokens through a pre-normalized cross-attention block with a residual
feed-forward projection, collapsing the clip into one pooled vector. A linear
layer with a softmax then produces the phase posterior, and the head is trained
with cross-entropy, optionally class-weighted and label-smoothed. Predictions
over successive clips form a phase timeline. Frame-F$_1$ measures class-balanced
recognition at the frame level, while phase mAP measures confidence-ranked
discrimination independently for each phase. To assess temporal structure,
F$_1$@10, F$_1$@25, and F$_1$@50 match predicted and reference segments at
increasing temporal-IoU thresholds, and the edit score measures agreement in
the ordered sequence of predicted phases while penalizing over-segmentation.

\textbf{Phase and step recognition (GraSP).}
For the longer-horizon GraSP tasks, we use the same lightweight temporal
segmentation head as for SAR-RARP50. The head consists of a learned single-query
spatial-attention pooler followed by one global temporal self-attention block
and a stack of residual dilated temporal convolutions. The design borrows
standard components from temporal action-segmentation architectures such as
MS-TCN and ASFormer~\cite{yi2021asformer}, but uses a simplified single-stage
topology. In our configuration the temporal sequence contains only 24 tokens,
making full global attention inexpensive and removing the computational
motivation for ASFormer's sliding-window attention.

The architecture is transferred from SAR-RARP50 to GraSP without structural
changes; only the output class count and class weights are changed. The learning
rate is selected from three schedules using a fold held out from the official
GraSP training cases, while the five official test cases are used only for the
reported score. For frozen evaluations, backbone features are cached per
surgery before training the head.

At each timestep, the spatial-attention pooler reduces the spatial token grid
to one descriptor. Fixed sinusoidal positional encoding and a single global
self-attention block then contextualize the temporal sequence, after which
residual dilated temporal-convolution blocks produce per-timestep class
predictions. The dilation increases across the stack to expand the temporal
receptive field. Training combines class-weighted cross-entropy with a
truncated temporal-smoothing (T-MSE) loss. We report mAP and F$_1$ separately
for phase and step recognition: mAP measures confidence-ranked class
discrimination, while F$_1$ measures class-wise prediction quality without
allowing the most frequent classes to dominate the evaluation.

\textbf{Action segmentation (SAR-RARP50).} The temporal SAR-RARP50 probe uses
the same spatial-attention-pooling and dilated-temporal-convolution head. The
encoder token grid is spatially pooled to a per-frame descriptor and processed
by the temporal stack to yield per-timestep action logits, trained with the same
weighted cross-entropy and temporal-smoothing objective and the same
multiple-learning-rate and validation-selection protocol. At test time,
overlapping clips are stitched into a per-video timeline. We report frame-macro
F$_1$ to measure class-balanced framewise recognition and segmental F$_1$@10,
which matches a predicted and reference action segment when their temporal IoU
is at least $0.10$. The latter penalizes missed and fragmented action segments
and is the principal metric used for comparison with the published
SAR-RARP50 benchmark.

\textbf{Surgeon-action detection (SARAS-ESAD).} The ESAD probe couples two heads
on the shared encoder features. A multi-label action-presence classifier, the
same temporal-segmentation head producing per-timestep multi-label logits
trained with a positive-weighted binary cross-entropy, predicts which actions
are present. In parallel, a class-conditioned bounding-box regressor attaches a
set of learnable per-class queries that cross-attend over the spatial token grid
at each timestep and regress normalized box coordinates through a small
feed-forward network, so that localization is conditioned on class identity by
construction. The box branch is trained only on present classes with a combined
$L_1$ and generalized-IoU loss. Detection AP is reported at box-IoU thresholds
of $0.10$, and $0.50$ (AP10, and AP50), with AP$_m$ denoting
their mean and therefore summarizing performance across localization
strictness. Presence mAP evaluates action recognition while ignoring
localization. Mean IoU and box mAP@50 instead isolate localization quality,
measuring geometric overlap and the fraction of correctly localized boxes,
respectively. As an exploratory analysis for this benchmark, we additionally evaluate our models with train-time data augmentation. We report results both with and without augmentation, ensuring that the direct comparison with the baselines, which are trained without augmentation, remains fair.


\section{Continued Pretraining at Production Scale}
\label{sec:pretraining}

The benchmark sections that follow treat each \cpt{} checkpoint as a fixed
encoder. This section reports how those checkpoints were produced, at what scale, and what practical trends emerged during pretraining. More broadly, we advocate for greater transparency in foundation-model pretraining for surgical robotics: downstream benchmark results should be accompanied by a detailed account of the data, training decisions, computational budget, and intermediate evidence that produced them.
The section is reported in full because its principal result also determines
the scope of every benchmark value that follows: within the configurations tested, the longest production run produced the strongest transfer, although its gain cannot be uniquely attributed to compute because its
catalogue also changed. At the budgets used to screen data composition, objective,
and sampling, surgical \cpt{} is statistically indistinguishable from the raw
initialization, so the null results obtained there are budget-scoped rather
than general, as are the frozen-probe ties reported later. Multi-seed effects
that exceed their own spread are distinguished throughout from single-seed
observations, which are reported as trends.

\begin{figure*}
    \centering
    \includegraphics[width=1.\linewidth]{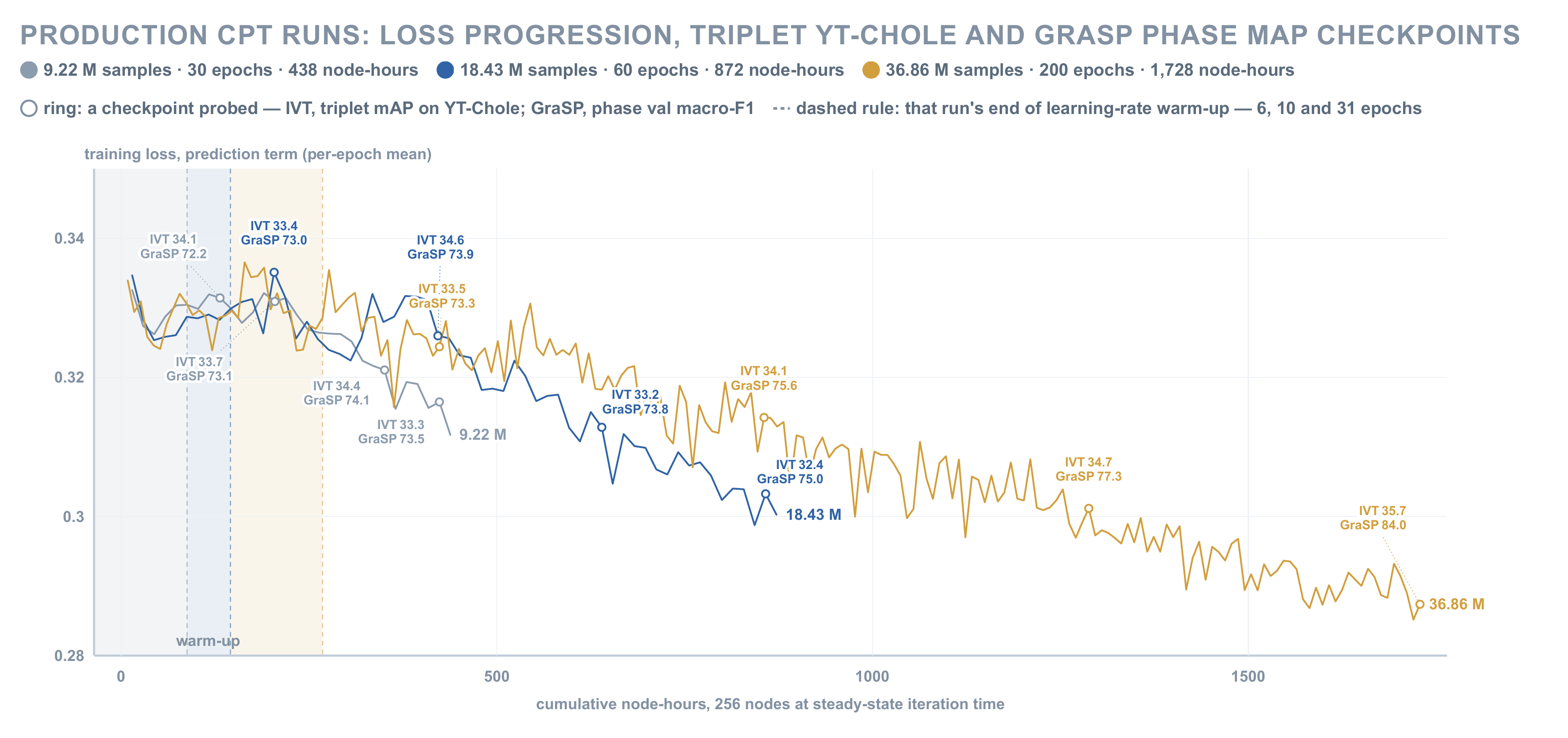}
\caption{\textbf{3,038 node-hours of surgical \cpt{}: scaling pretraining and downstream transfer.}
Each curve shows the prediction-term training loss for one of three production \cpt{} runs with increasing pretraining budgets; circled points mark checkpoints at which we probe the learned representation on IVT recognition (triplet mAP on YT-Chole) and GraSP phase recognition (validation macro-F$_1$). Shaded regions extend to the end of each run's learning-rate warm-up. Across successively larger budgets, optimization continues to improve while the downstream probes show complementary transfer behavior: the smaller runs exhibit task-dependent saturation and checkpoint sensitivity, whereas the 36.86,M-sample run combines the lowest training loss with continued improvement in both probes, reaching 35.7 IVT mAP and 84.0 GraSP phase macro-F$_1$ at its final probe. The resulting 36.86,M encoder is used for our main model and subsequently reaches 85.34 GraSP phase mAP after last-eight-block fine-tuning, $+8.6$ points over the best published system (Fig.~\ref{fig:grasp-bars}).}
    \label{fig:pretraining}
\end{figure*}
\subsection{Campaign scale and setup}

Continued pretraining resumes the released V-JEPA-2.1 objective on a
multi-source surgical catalogue at $384^2$ resolution and 16 frames. The catalogue
used for the reported 2\,B production runs is a 14-entry surgical catalogue representing 19 distinct datasets (the six small classic sets listed below enter as a single
shard) and approximately 2,650 hours of video, mixed by a
temperature-weighted sampler over per-source clip counts. Hours were estimated
from the 60-second-segment counts of the packed web shards rather than from the
raw source files, so they reflect the material the loader receives; because not
every source is packed at a uniform 60-second segment length, these hour figures
are approximate.

By hours, the reported catalogue is roughly 51\% robotic (our private corpus, together with SurgVU24, SurgToolLoc2022, SurgeNet-robotic, and the leak-free GraSP shard)~\cite{zia2025surgvu,zia2023surgtoolloc,jaspers2026surgenet,ayobi2025grasp}; about one third comes from a single large mixed-surgical source, the YouTube-scraped LEMON corpus~\cite{che2026lemon}; roughly 15\% is purely laparoscopic (Cholec80, MultiBypass140, SurgeNet-laparoscopic, GynSurg, LapGyn6-events, and HeiChole)~\cite{twinanda2017endonet,lavanchy2024multibypass,jaspers2026surgenet,nasirihaghighi2025gynsurg,nasirihaghighi2024lapgyn6events,wagner2023heichole}; and a small remainder comes from six classic sets (CRCD, EndoVis15, JIGSAWS, MICCAI-2017, MICCAI-EndoSeg, and SurgVisDom)~\cite{oh2025crcd,bodenstedt2018endovis15,gao2014jigsaws,allan2019endovis17,allan2020endovis18,zia2021surgvisdom}. The catalogue composition differs across the three production runs. The 9.22,M and 18.43,M runs use a surgical-only catalogue with no general or natural-video anchor. The 36.86,M run, which supplies the headline encoder, additionally restores the Kinetics-400 and Open-H-Embodiment shards as rehearsal anchors~\cite{kay2017kinetics,openh2026embodiment}, for the reasons given below. Robotic da Vinci video constitutes the majority of the surgical mixture by hours in all three runs. Robotic da Vinci video therefore constitutes the majority of the
mixture by hours, and this is the composition under which the benchmark
results of this paper were obtained.

Contamination is prevented before pretraining rather than audited after it.
Every YouTube-scraped source passes a crop- and mask-augmented perceptual-hash
gate: sampled frames of each candidate training video are matched against the
evaluation frames of the probe benchmarks, and any source video with more than
10\% of its frames within a Hamming distance of 6/64 of an evaluation frame is
flagged as leaked and dropped from the catalogue. The augmentation is what makes
the gate usable on scraped video, where re-uploads are routinely cropped,
letterboxed, or logo-masked and defeat a plain hash. GraSP is excluded by
identity instead: its shard was resharded with the five official test cases
(041, 047, 050, 051, 053) removed by case ID. The remaining sources are
institutional or benchmark-release corpora, disjoint from the evaluation splits
by provenance.

In total the campaign comprised more than 3 continued-pretraining runs and
more than 254 downstream probe and fine-tuning runs, 109 of which unfreeze the
last blocks of the backbone rather than probing it frozen.

Production runs use 256 compute nodes with a global batch of 6{,}144 and a
fixed 50 iterations per epoch, so epoch counts are directly comparable across
arms of the same run. Three production runs are reported. The first two, a
9.22\,M-sample run and a $2\times$-budget 18.43\,M-sample run, are identical in
catalogue and recipe and differ only in samples seen; they are the pair that
carries the production budget comparison. The third doubles the budget again
to 36.86\,M samples but is \emph{not} a fourth point on that curve: it restored the Kinetics-400 and Open-H-Embodiment shards in response to general-video forgetting, measured on Something-Something-v2 (SSv2) by validation top-1 accuracy and macro-F1 relative to the raw Meta 2B initialization. Fixing the iteration count
per epoch rather than the samples per epoch was a deliberate bookkeeping
choice: at this node count the effective epoch length otherwise drifts with
loader throughput and stragglers, and checkpoint indices stop being comparable
between arms. This convention must be fixed before the first launch, since it
cannot be reconstructed afterwards from the checkpoints alone.

The single-variable comparisons could not be run at production scale. Smaller 1 B ViT-g screening runs were used to explore composition, objective, and sampling choices before the production campaign. This two-tier structure, a small number of expensive
production runs together with a larger set of inexpensive screening arms, is
the practical form such a campaign takes, and the budget control in
the following subsection determines the extent to which the screening results
generalize.

\subsection{Transfer across production-scale runs}

The principal scaling lever in our \cpt{} pre-training was the amount of
production compute devoted to the encoder. Figure~\ref{fig:pretraining}
summarises the three runs that led to the final model: a 9.22\,M-sample run
trained for 438 node-hours, an 18.43\,M-sample run trained for 872 node-hours,
and the main 36.86\,M-sample run trained for 1,728 node-hours, for a total of
3,038 node-hours. Rather than treating these runs only as optimisation
traces, we periodically evaluated checkpoints with two lightweight downstream
probes. We selected IVT recognition on YT-Chole and phase recognition on GraSP
because they responded differently to continued pre-training and therefore
provided complementary views of representation quality.

Importantly, these probes are diagnostic rather than optimisation objectives.
Their labels and losses never enter the pre-training update rule, which is
driven exclusively by the self-supervised $\ell_1$ prediction loss; their
scores are used only as development-time diagnostics. They therefore provide
read-only measurements of whether reductions in the pre-training objective are
accompanied by increasingly transferable surgical features. This distinction
is precisely why they are useful alongside training loss: the latter measures
progress on the pre-training objective, whereas the probes test whether that
progress is reflected in downstream representation quality.

The first two runs already illustrate why this distinction matters. In the
9.22\,M-sample run, IVT follows a non-monotonic trajectory of 34.1, 33.7, 34.4,
and 33.3 mAP, while GraSP progresses from 72.2 to 73.1, 74.1, and 73.5. The
18.43\,M-sample run behaves similarly: IVT rises from 33.4 to 34.6 before
falling to 33.2 and 32.4, whereas GraSP remains more stable at 73.0, 73.9,
73.8, and 75.0. Thus, continued reduction of the self-supervised objective does
not uniquely determine the trajectory of downstream transfer, and the two
probes respond differently to the same optimisation process.

The largest run changes this picture. With the 36.86\,M-sample budget, both
downstream measurements improve over the longer optimisation horizon. IVT
progresses from 33.5 to 34.1, 34.7, and finally 35.7 mAP, while GraSP moves from
73.3 to 75.6, 77.3, and ultimately 84.0. The important observation is therefore
not that every probe improves uniformly at every point in training, but that
the largest production run eventually enters a regime in which continued
self-supervised optimisation is accompanied by clear gains in both transfer
metrics. GraSP provides the strongest evidence of this transition, while IVT
shows that it extends to a distinct downstream task.

The largest production configuration consequently yields the strongest
transfer, but its catalogue change prevents attributing this gain to budget
alone. We progressively increased the production budget from 9.22\,M to
18.43\,M and ultimately 36.86\,M samples, culminating, to our knowledge, in one
of the largest pre-training campaigns reported for robotic-assisted surgery
(3,038 node-hours). Intermediate checkpoints were evaluated with IVT and
GraSP not to optimise their task metrics, but to verify whether additional
self-supervised optimisation continued to produce increasingly transferable
representations. The final 36.86\,M-sample run simultaneously achieves the
lowest pre-training loss and the strongest probe performance, and its encoder
forms the basis of our main experiments, subsequently reaching 85.34 phase mAP
on GraSP.

We nevertheless separate this observation from a causal claim about compute.
Under the matched catalogue and recipe, increasing the budget from 9.22\,M to
18.43\,M only modestly improves the final GraSP probe from 73.5 to 75.0, while
IVT decreases from 33.3 to 32.4. The clearest transfer gain therefore, coincides
with the 36.86\,M-sample run, whose final probes reach 35.7 IVT mAP and 84.0
GraSP, but for which both the budget and the catalogue changed. We conclude that the
longest production run produced the strongest representation among the
configurations tested, without attributing that improvement to compute alone.

\subsection{Lessons from a 256-node campaign}

The campaign suggests a practical protocol for training large-scale surgical video representations for robotic-assisted surgery. Its central principle is to protect sustained pretraining duration while monitoring representation quality throughout training with a small set of complementary downstream probes. We use IVT recognition on YT-Chole and phase recognition on GraSP because they place different demands on the representation. IVT operates at a relatively short temporal horizon and requires spatially and semantically grounded recognition of instruments and their interactions with the surgical scene, whereas phase recognition depends on longer-range procedural context and progression. The two probes therefore test complementary aspects of surgical representation quality rather than redundant readouts of the same capability.

This complementarity is also visible empirically as pretraining progresses. IVT is sensitive to checkpoint choice and exposes regressions that are not apparent from the pretraining objective alone, whereas GraSP more clearly resolves gains that emerge only after sustained large-budget training. Their disagreement at intermediate checkpoints is therefore useful rather than problematic: it shows that continued improvement of the self-supervised objective does not translate uniformly across downstream capabilities. Evaluating both at fixed checkpoints provides a richer picture of representation development than either training loss or a single downstream task in isolation. Importantly, these probes remain read-only diagnostics: their labels and metrics never enter the pretraining update rule, which is driven exclusively by the self-supervised objective.

Within this campaign, the clearest transfer gains emerge in the final production configuration. Smaller screening runs did not resolve consistent effects of catalogue composition, sampling temperature, or masking objective. The practical implication is not that these factors are universally unimportant, but that small-budget screening can fail to reveal effects that become visible only once the model is trained substantially longer. In our campaign, allocating compute to the final production configuration proved more useful than further branching the smaller screening runs.

This interpretation nevertheless requires an important qualification. We do not claim that compute alone caused the final improvement. The only matched budget comparison, from 9.22,M to 18.43,M samples under the same catalogue and recipe, was approximately probe-neutral, whereas the 36.86,M-sample run that produced the clearest transfer gain also changed its training catalogue. The production series therefore supports a recommendation about where additional effort proved most useful in our campaign, rather than a controlled scaling law attributing improvement uniquely to training budget.

Two additional diagnostics further qualify the scaling picture. Measured without a task head, the production checkpoint reconstructs masked surgical video more accurately than its initialization across every independent unit of five corpora, showing that frozen-probe ties need not imply an unchanged representation. At the same time, on SSv2, the 36.86M-sample checkpoint drops from 60.27 to 53.82 top-1 accuracy and from 52.58 to 44.70 macro-F1 relative to the raw Meta 2B initialization (-6.45 and -7.88 points, respectively). Large-scale surgical \cpt{} should therefore be viewed as specialization: improvements in target-domain representation quality can coexist with degradation outside the surgical domain, and this trade-off is not visible from the surgical benchmark suite alone.

Together, these observations motivate three operational recommendations for comparable robotic-assisted-surgery pretraining campaigns.

\emph{Use complementary probes throughout production training.}
Select probes with different temporal and semantic requirements in advance and evaluate them at fixed checkpoints. Here, YT-Chole IVT captures short-horizon, spatially grounded surgical activity, while GraSP phase recognition probes longer-horizon procedural context. Their joint trajectory reveals both checkpoint sensitivity and whether continued self-supervised optimization is producing increasingly transferable representations.

\emph{Allocate budget to long runs before allocating it to variants.}
Clear transfer gains emerged only beyond inexpensive screening budgets. A null result at small scale should therefore not be interpreted as strong evidence that an intervention is ineffective. Once the recipe and evaluation harness are validated, our results favor extending a strong configuration to production scale before distributing the same compute across many shallow variants.

\emph{Lock reproducibility conventions and diagnostic probes before launch.}
Checkpoint indexing, epoch definitions, data manifests, and probe locations should be fixed as part of the training specification, since ambiguities are difficult to reconstruct after long distributed runs. The evaluation suite should also include at least one diagnostic outside the target surgical benchmarks: our general-action forgetting probe was the only measurement that moved against the otherwise favorable surgical trend, exposing a trade-off that the surgical tasks alone could not reveal.

The resulting protocol is intentionally simple: validate the harness, choose probes that span distinct representation requirements, run sufficiently long production training, and evaluate fixed checkpoints throughout. For robotic-assisted surgery, this turns pretraining from a single end-point experiment into a monitored scaling process in which training loss, downstream transfer, and out-of-domain retention provide complementary evidence about what the representation is actually learning.

\section{Results}
\label{sec:results}

\begin{figure}[tb]
\centering
\includegraphics[width=\linewidth]{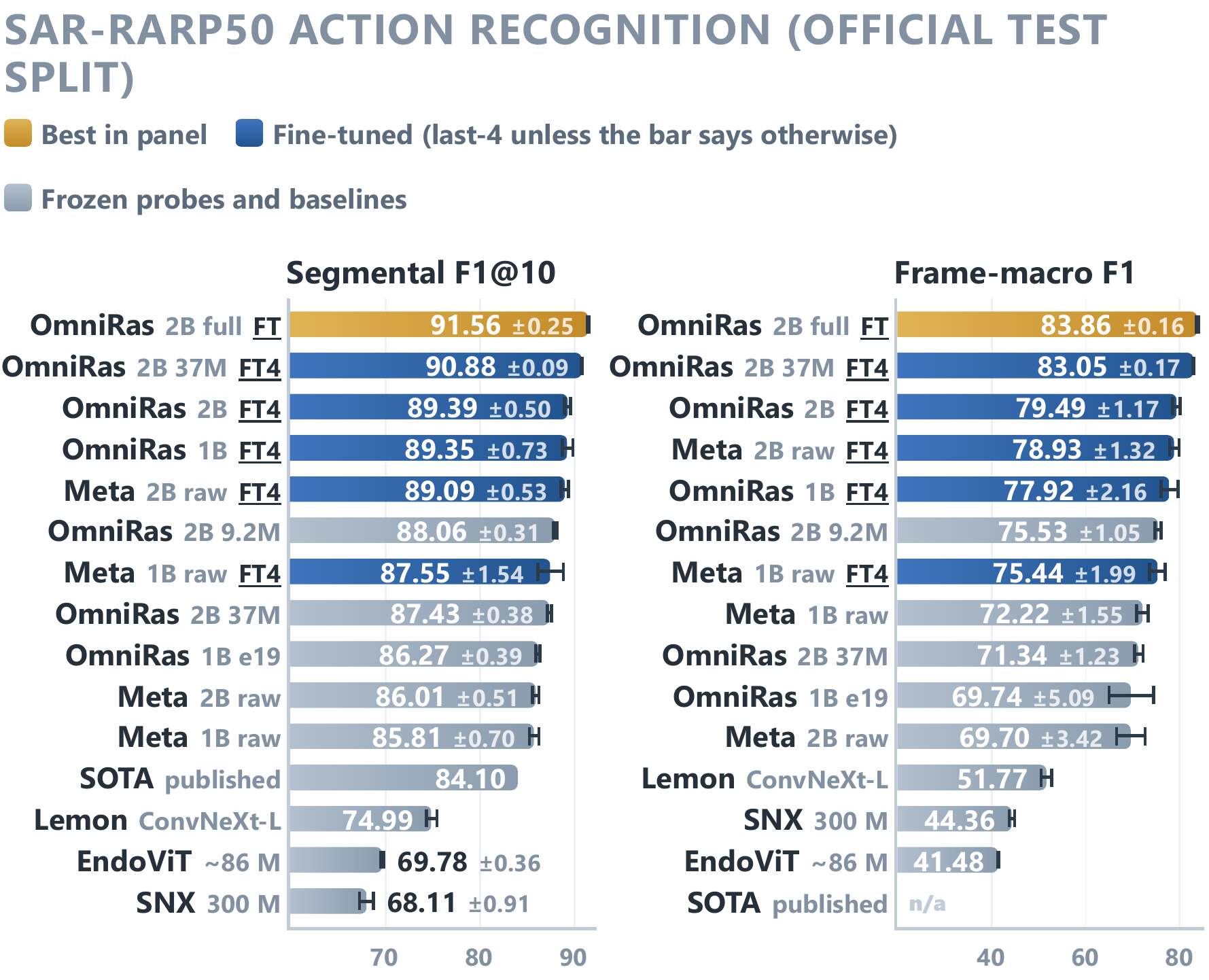}
\caption{\textbf{SAR-RARP50 official-test action recognition} using segmental
F$_1$@10 and per-frame macro-F$_1$; error bars show standard deviation over
three seeds where available.
}
\label{fig:sar-bars}
\end{figure}

\begin{figure*}[!tb]
\centering
\includegraphics[width=\linewidth]{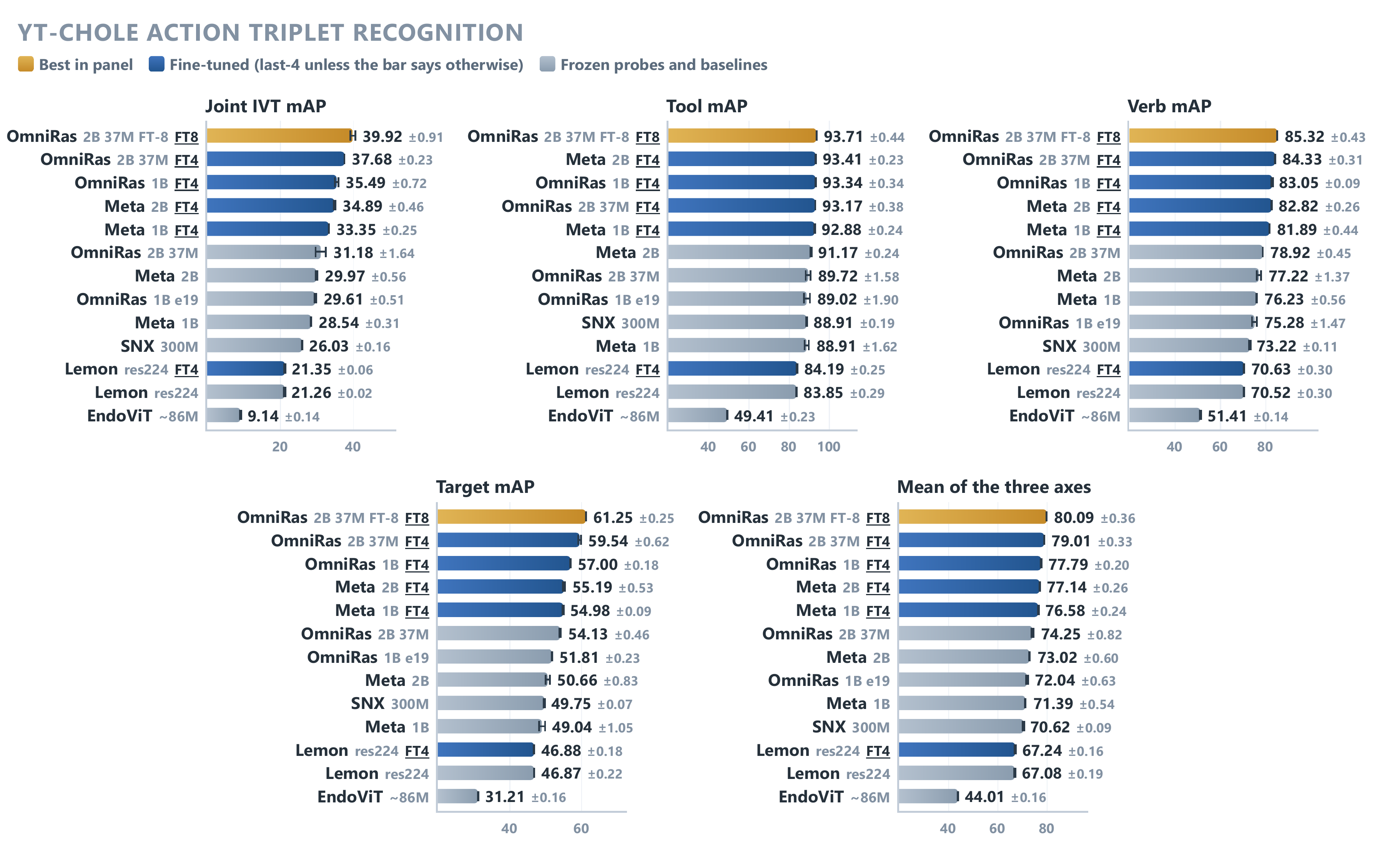}
\caption{\textbf{YT-Chole action-triplet recognition probe.} Spans across tool, verb, target, and
mean and joint IVT metrics. Error bars show standard deviation over three
seeds.
}
\label{fig:triplet-bars}
\end{figure*}
We evaluate the representations through 254 downstream seed-level runs, of which 109 involve partial fine-tuning of the backbone. This evaluation budget was chosen not only to cover multiple datasets and transfer settings, but also to quantify the variability of downstream adaptation: wherever applicable, we therefore report mean$\pm$std over three independent seeds rather than relying on a single run. This distinction is consequential in the present setting. Among the ten matched raw-versus-\cpt{} comparisons summarized below, three have overlapping $\pm1$ standard-deviation intervals, showing that differences between encoders can be comparable to the variation introduced by downstream training itself. Multi-seed evaluation is therefore essential for separating reproducible transfer effects from fluctuations of a particular optimization run.

Unless otherwise stated, each checkpoint is evaluated as a fixed encoder or under the specified 4-layers partial fine-tuning protocol. 
We use frozen and FT4 results for matched raw-versus-CPT comparisons; FT8, full fine-tuning, and augmentation are reported only as task-specific best configurations.

We intentionally present frozen and FT4 regimes together because they answer complementary questions about the same initialization. Frozen evaluation measures directly accessible representation quality, whereas FT4 tests whether continued pretraining provides a stronger predictive prior for adaptation to surgical video than the generic seed. Keeping both views together therefore distinguishes gains in representation accessibility from gains that emerge under a controlled, matched adaptation budget.

\subsection{YT-Chole action triplets}

Figure~\ref{fig:triplet-bars} summarizes action-triplet recognition on YT-Chole. Across frozen encoders, the billion-scale video models substantially outperform the smaller surgical baselines on joint IVT mAP. The strongest frozen result is obtained by the longer-trained 37M-sample OmniRAS 2B checkpoint, reaching $31.18\pm1.64$ IVT, compared with $29.97\pm0.56$ for the raw 2B initialization and $29.61\pm0.51$ for OmniRAS 1B. At matched pretraining budgets, however, the frozen effect is mixed: OmniRAS changes IVT by $+1.07$ at 1B and $-0.44$ at 2B relative to the corresponding raw encoders. Thus, surgical pretraining alone does not yield a uniform advantage under a strictly frozen probe.

The picture changes once limited backbone adaptation is allowed. With FT4, OmniRAS improves over the corresponding raw encoder at both model scales, from $33.35\pm0.25$ to $35.49\pm0.72$ IVT at 1B and from $34.89\pm0.46$ to $36.09\pm0.25$ at 2B. Extending OmniRAS pretraining to 37M samples further raises the 2B result to $37.68\pm0.23$ under FT4, while FT8 reaches the overall best performance of $39.92\pm0.91$. The improvement is also visible in the triplet components: the 37M FT8 model reaches $93.71$ tool, $85.32$ verb, and $61.25$ target AP, with target recognition showing the largest remaining headroom. These results suggest that the benefit of OmniRAS is expressed most consistently when the pretrained representation can be partially adapted to the downstream task, rather than when it is evaluated as a completely fixed encoder.

\subsection{OmniRAS-PR phase recognition}

\begin{figure*}[tb]
\centering
\includegraphics[width=\linewidth]{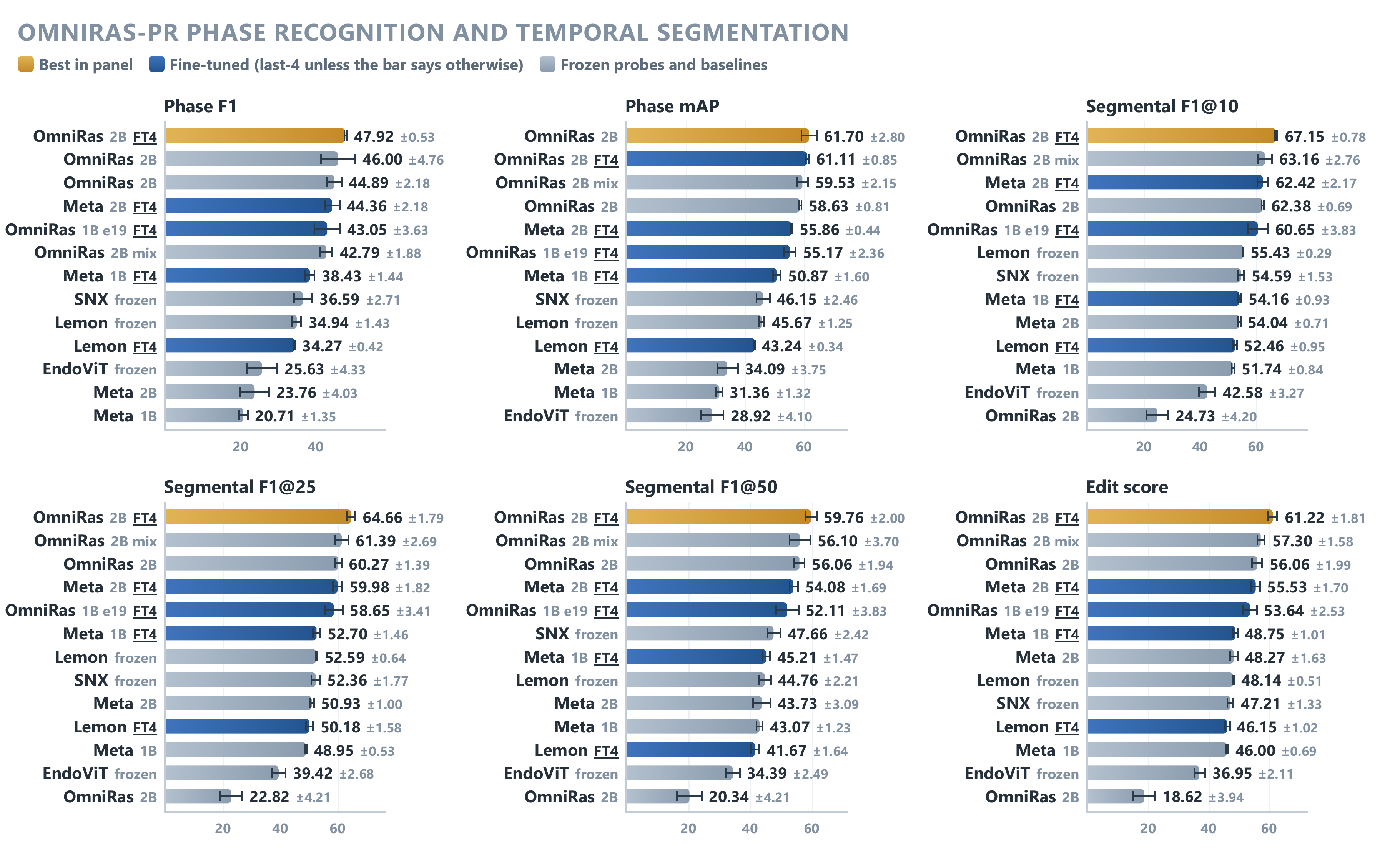}
\caption{\textbf{OmniRAS-PR} frame-wise phase-recognition and temporal-segmentation
metrics. Error bars show the three-seed standard deviation. 
}
\label{fig:ytphase-bars}
\end{figure*}

Figure~\ref{fig:ytphase-bars} evaluates both frame-wise phase recognition and temporal segmentation. Frame-F$_1$ and phase mAP measure per-frame discrimination, whereas F$_1$@10/25/50 and Edit additionally penalize fragmented and over-segmented phase sequences. OmniRAS 2B with last-four-block adaptation reaches $47.92\pm.53$ frame-F$_1$, $61.11\pm.85$ phase mAP, and $67.15\pm.78$ F$_1$@10, performing strongly across the reported metrics. More generally, the adapted configurations occupy the top of the frame-level ranking, indicating that limited backbone adaptation is consistently beneficial for this task.

The frozen comparison shows a strong effect of surgical pretraining. Relative to the raw Meta 2B encoder, frozen OmniRAS 2B improves frame-F$_1$ from $23.76\pm4.03$ to $44.89\pm2.18$ and phase mAP from $34.09\pm3.75$ to $58.63\pm.81$. Under matched last-four-block adaptation, OmniRAS 2B reaches $47.92\pm.53$ frame-F$_1$ and $61.11\pm.85$ phase mAP, compared with $44.36\pm2.18$ and $55.86\pm.44$ for Meta 2B FT4. OmniRAS also leads the segmental metrics under FT4, reaching $67.15\pm.78$ F$_1$@10, $64.66\pm1.79$ F$_1$@25, $59.76\pm2.00$ F$_1$@50, and $61.22\pm1.81$ Edit.

The frozen OmniRAS 2B representation also exceeds SurgeNetXL on both frame-wise and temporal metrics, including $44.89$ versus $36.59$ frame-F$_1$ and $62.38$ versus $54.59$ F$_1$@10. Partial adaptation further improves temporal consistency, with OmniRAS 2B FT4 reaching $67.15$ F$_1$@10, $59.76$ F$_1$@50, and $61.22$ Edit.

\subsection{SAR-RARP50 action recognition}

\begin{figure*}[tb]
\centering
\includegraphics[width=\linewidth]{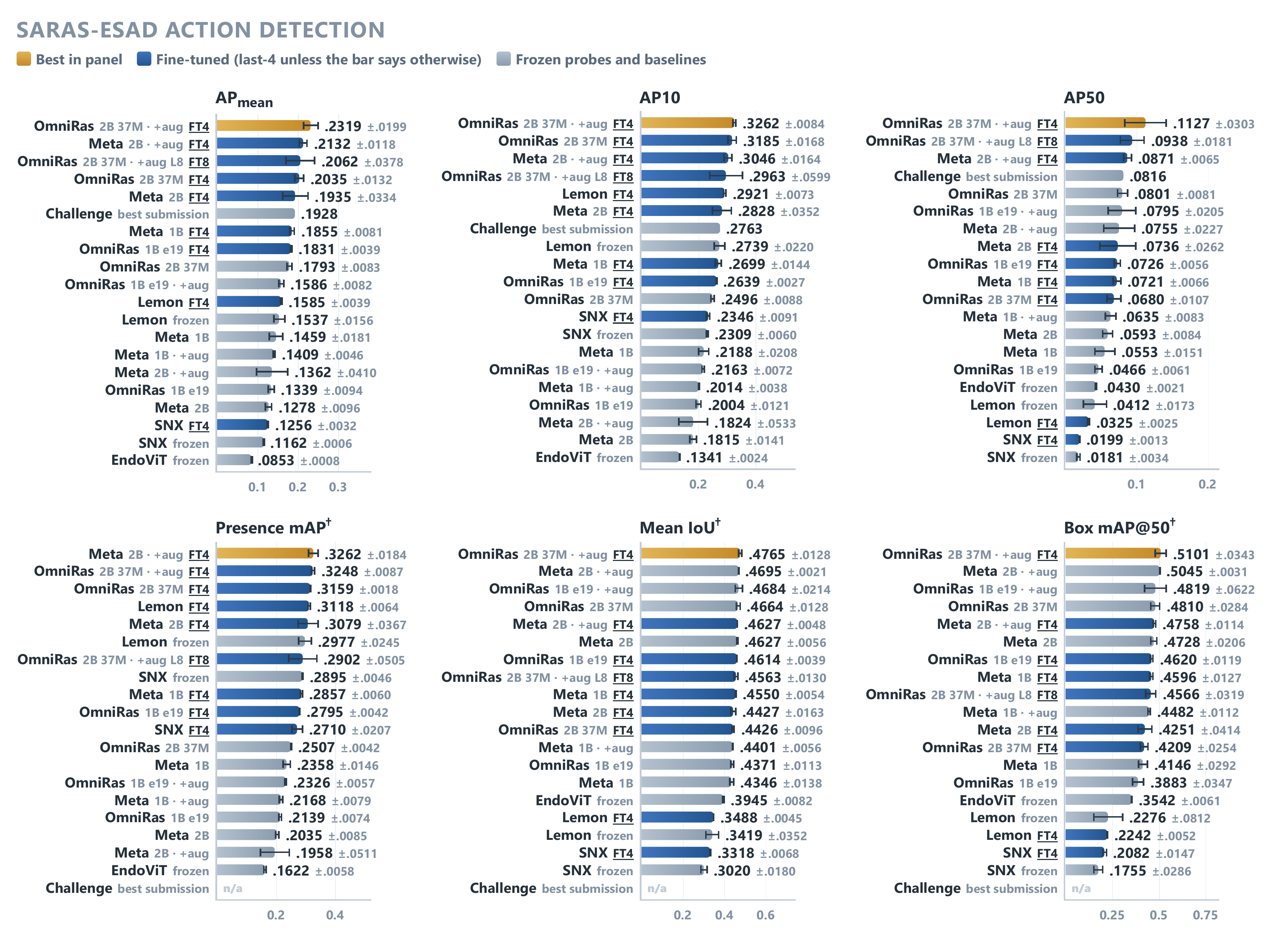}
\caption{S\textbf{ARAS-ESAD detection} AP and diagnostic recognition/localization
scores for frozen and last-four-block-adapted encoders. Error bars show standard deviation over three seeds.
}
\label{fig:esad-bars}
\end{figure*}
The controlled comparison in Figure~\ref{fig:sar-bars} uses the official SAR-RARP50 test split. Under frozen probing, the billion-scale video encoders form a tight high-performing group, all above the published $84.10$ F$_1$@10 reference. OmniRAS improves over the corresponding Meta initialization at both scales, from $85.81\pm0.70$ to $86.27\pm0.39$ at 1B and from $86.01\pm0.51$ to $87.40\pm0.38$ at 2B. The longer-trained OmniRAS checkpoints remain in the same range, reaching $88.06\pm0.31$ after 9.2M samples and $87.43\pm0.38$ after 37M samples. These results show that strong action representations are already accessible without adapting the backbone.

Partial adaptation raises the ceiling further. With the last four blocks unfrozen, OmniRAS reaches $89.35\pm0.73$ at 1B and $89.39\pm0.50$ at 2B, compared with $87.55\pm1.54$ and $89.09\pm0.53$ for the corresponding Meta models. The 37M OmniRAS 2B checkpoint reaches the strongest result in the table at $91.56\pm 0.25$ F$_1$@10 and $83.86\pm0.16$ frame-macro accuracy when fully unfreezed. Relative to the published $84.10$ reference, this is a 7.46-point gain in segmental F$_1$@10.

The two metrics also expose different aspects of transfer. Frozen F$_1$@10 varies only from $85.81$ to $88.06$ across the billion-scale video encoders, whereas frame-macro accuracy is considerably less ordered: for example, Meta 1B reaches $72.22\pm1.55$ while OmniRAS 1B obtains $69.74\pm5.09$. After adaptation, this discrepancy narrows and OmniRAS 37M 2B leads both metrics. The result suggests that the main benefit of adapting OmniRAS is not simply better frame-wise discrimination, but improved access to the temporal structure required by segmental action recognition.

\subsection{SARAS-ESAD action detection}

Figure~\ref{fig:esad-bars} summarizes SARAS-ESAD action detection. OmniRAS 2B
trained on 37M samples and adapted through the last four blocks gives the
strongest overall performance, reaching $0.2035\pm0.0132$ AP$_\text{mean}$.
Adding augmentation further increases this to $0.2319\pm0.0199$, with
corresponding gains to $0.3262\pm0.0084$ AP10, and $0.1127\pm0.0303$ AP50. This
configuration exceeds the best challenge submission, which obtains $0.1928$
AP$_\text{mean}$, and achieves the highest value in every detection AP metric.

The results show a consistent benefit from downstream adaptation of OmniRAS. For
the 37M 2B checkpoint, FT4 increases AP$_\text{mean}$ from $0.1793\pm0.0083$ when
frozen to $0.2035\pm0.0132$. Augmentation, which we apply to our models only,
adds a further $+0.028$ AP$_\text{mean}$ on top of FT4 (to $0.2319\pm0.0199$) and
consistently helps across our configurations,for example lifting the 2B FT4
result by a comparable margin, indicating headroom on this benchmark that the
no-augmentation baselines do not yet exploit. The earlier OmniRAS 2B checkpoint
follows the same pattern, rising from $0.1395\pm0.0168$ when frozen to
$0.2005\pm0.0080$ with FT4. At 1\,B, adaptation similarly raises OmniRAS to
$0.1831\pm0.0039$, placing it among the strongest configurations in the table.

The localization diagnostics reveal a different trend from detection AP.
Adaptation improves action detection without necessarily improving box geometry:
for the 37M OmniRAS 2B model, mean IoU decreases from $0.4664\pm0.0128$ to
$0.4426\pm0.0096$ after FT4, while box mAP@50 decreases from $0.4810\pm0.0284$ to
$0.4209\pm0.0254$. Conversely, the frozen OmniRAS 2B checkpoint obtains the
highest localization diagnostics, reaching $0.4822\pm0.0125$ mean IoU and
$0.5156\pm0.0275$ box mAP@50 despite lower detection AP. The gain from adaptation
therefore appears to arise primarily from improved action discrimination and
confidence ranking rather than from uniformly more accurate localization.

\subsection{GraSP phase and step recognition}

\begin{figure}[!t]
\centering
\includegraphics[width=\linewidth]{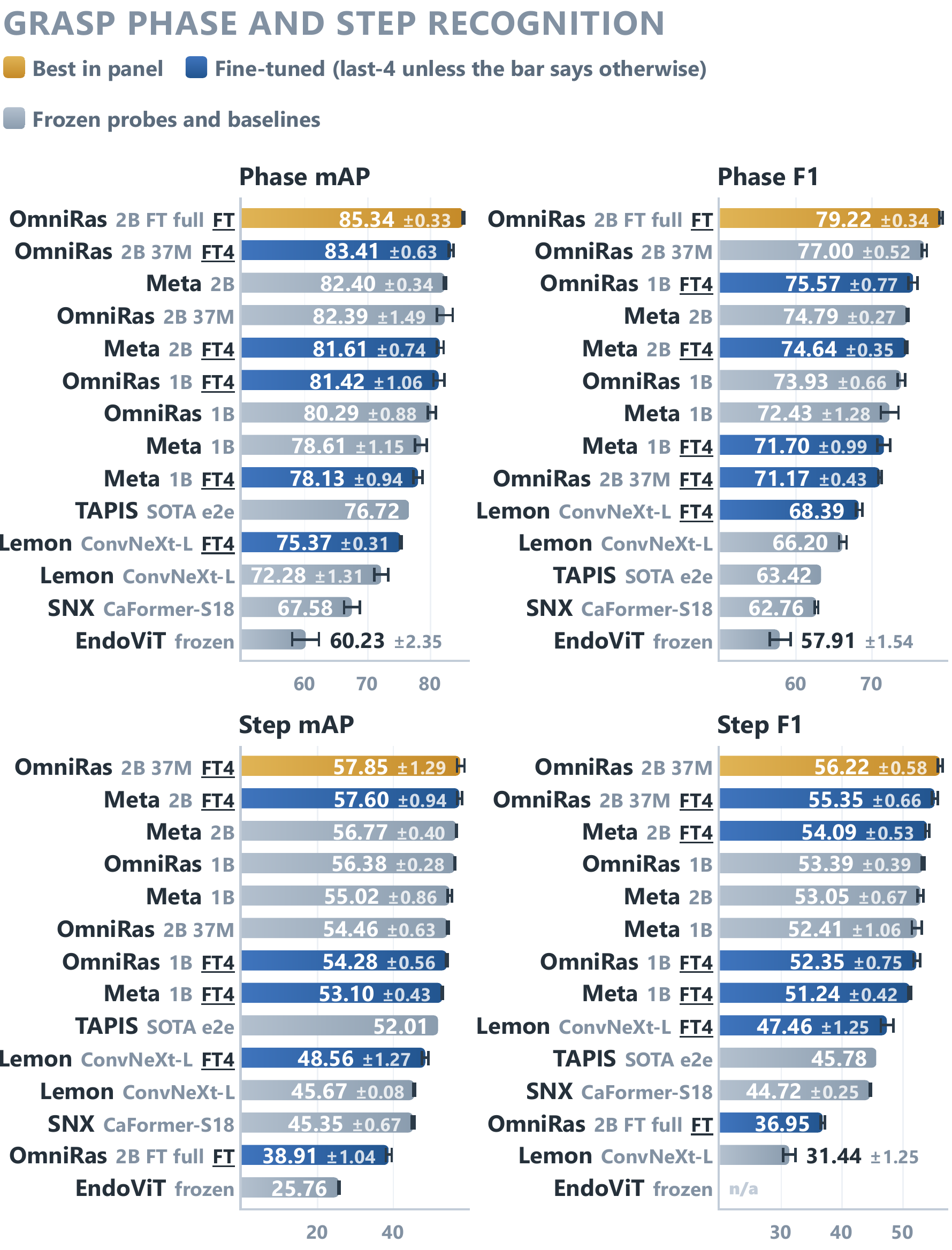}
\caption{\textbf{GraSP phase- and step-recognition} mAP and macro-F$_1$. Exact
three-seed results are shown with standard-deviation error bars.
}
\label{fig:grasp-bars}
\end{figure}

Figure~\ref{fig:grasp-bars} reports phase and step recognition jointly. GraSP is evaluated as an out-of-the-box transfer of the SAR-RARP50 head, without task-specific tuning on the GraSP test set. Across the table, the billion-scale video encoders consistently outperform the specialized image baselines, while most also exceed TAPIS on both phase and step mAP.

Under frozen probing, Meta 2B and the 37M OmniRAS 2B checkpoint obtain nearly identical phase mAP, at $82.40\pm0.34$ and $82.39\pm1.49$, respectively, while Meta 2B is stronger on step mAP, at $56.77\pm0.40$ versus $54.46\pm0.63$. After last-four-block adaptation, the ordering favors OmniRAS: the 37M OmniRAS 2B FT4 model reaches $83.41\pm0.63$ phase mAP and $57.85\pm1.29$ step mAP, compared with $81.61\pm0.74$ and $57.60\pm0.94$ for Meta 2B FT4.

GraSP also exposes a clear trade-off between coarse phase recognition and finer-grained step recognition. Full fine-tuning of OmniRAS 2B produces the highest phase result in the table, $85.34\pm0.33$ mAP, but step mAP falls sharply to $38.91\pm1.04$. Restricting adaptation to the last four blocks preserves substantially stronger step recognition, reaching $57.85\pm1.29$ step mAP while maintaining $83.41\pm0.63$ phase mAP. This divergence suggests that aggressive adaptation can specialize the representation toward coarse workflow phases at the expense of the finer temporal distinctions required for step recognition.

The same adaptation effect is visible for LemonFM, whose FT4 variant improves from $72.28$ to $75.37\pm0.31$ phase mAP and from $45.67$ to $48.56\pm1.27$ step mAP. Nevertheless, the adapted image baseline remains below the strongest video encoders, particularly on step recognition. Taken together, the results indicate that limited backbone adaptation is important for GraSP transfer, but that preserving a balance between phase-level and step-level information is more important than maximizing either task in isolation.

\subsection{Cross-task synthesis}
\begin{figure*}[tb]
\centering
\includegraphics[width=\linewidth]{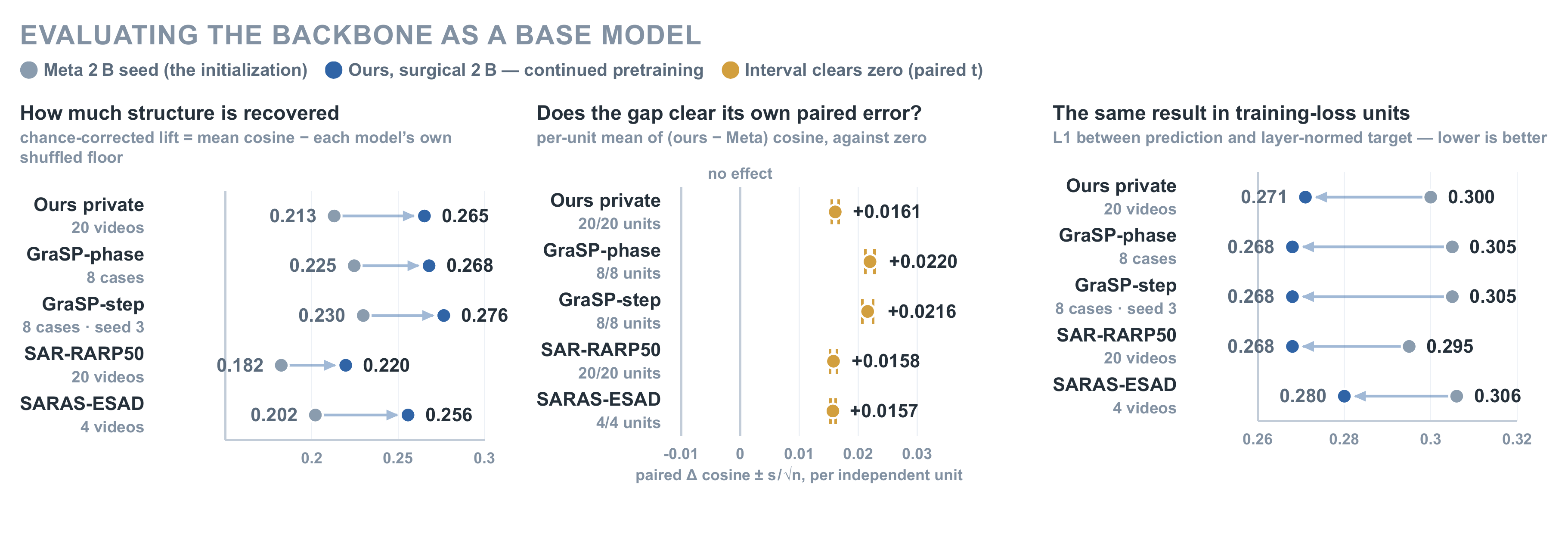}
\caption{\textbf{Masked-prediction fidelity of the 2B surgical checkpoint} against the
Meta 2B checkpoint it was pretrained from, measured by replicating the
V-JEPA-2.1 objective with no task head. Left: chance-corrected lift, each
model's mean cosine minus its own shuffled-target floor. Center: the paired
per-unit difference in mean cosine against zero; intervals are
$s_d/\sqrt{n}$ over independent source videos or surgical cases, and are
smaller than the plotted dot at this scale. Right: L1 between prediction and
layer-normed target, the actual training loss, where lower is better. GraSP-step
re-samples the same corpus as GraSP-phase under a second seed and is a
robustness check rather than an independent benchmark.}
\label{fig:pretrain-fidelity}
\end{figure*}
Across the downstream suite, OmniRAS reaches the strongest result in every surgical task family evaluated. On YT-Chole triplets, the 37M-sample OmniRAS 2B model reaches $39.92\pm0.91$ IVT mAP with FT8. On OmniRAS-PR phase recognition, OmniRAS 2B FT4 achieves $47.92\pm.53$ frame-F$_1$, $61.11\pm.85$ phase mAP, and $67.15\pm.78$ F$_1$@10. On SAR-RARP50, the 37M model reaches $90.88\pm0.09$ F$1$@10 and $83.05\pm0.17$ frame-macro accuracy, above the published $84.10$ F$1$@10 reference. The same model family also gives the strongest SARAS-ESAD result: OmniRAS 2B FT4 with augmentation reaches $0.2319\pm0.0199$ AP$\text{mean}$ and leads AP10, and AP50, compared with $0.1928$ AP$\text{mean}$ for the best challenge submission.

GraSP evaluation shows that this advantage extends beyond a single level of surgical workflow recognition. On GraSP, the 37M OmniRAS 2B FT4 model attains $83.41\pm0.63$ phase mAP and $57.85\pm1.29$ step mAP, improving over TAPIS by 6.69 and 5.84 points, respectively. The task-specific optima occur under different adaptation regimes: full fine-tuning achieves the highest phase-recognition mAP at $85.34\pm0.33$, whereas last-four-block fine-tuning achieves the strongest step-recognition mAP at $57.85\pm1.29$. This distinction is important. The strongest transfer does not come from applying the same adaptation depth everywhere, but from adjusting how much of the pretrained representation is updated for the downstream task.

A second pattern is that OmniRAS benefits are most consistent once part of the backbone is allowed to adapt. Frozen comparisons are more mixed, whereas partial fine-tuning more clearly exposes the advantage of surgical pretraining: it improves over the corresponding raw initialization on both OmniRAS-PR scales, reverses the frozen ordering on GraSP, and produces the strongest OmniRAS-PR phase-recognition and SARAS-ESAD models. Longer surgical pretraining adds another gain in several settings: on YT-Chole triplets, the 37M checkpoint improves the 2B FT4 result from $36.09\pm0.25$ to $37.68\pm0.23$, and FT8 raises it further to $39.92\pm0.91$.

Importantly, this specialization does not require discarding the general-video representation. With Kinetics-400 rehearsal during surgical pretraining, OmniRAS reaches $53.82$ top-1 and $44.70$ macro-F1 on SSv2, compared with $60.27$ and $52.58$ for the raw 2B initialization. Taken together, the 254 downstream seed-level runs show that OmniRAS is not a task-specific gain: it provides a strong surgical-video initialization across recognition, temporal segmentation, detection, and workflow transfer while retaining most of the starting model's general-video capability.

\subsection{Label-free evaluation of objective-aligned predictive fidelity}
\label{sec:fidelity}

\begin{table}[!h]
\centering
\caption{Masked-prediction fidelity, 20 clips per benchmark per model.
}
\label{tab:pred-fidelity}
\renewcommand{\arraystretch}{1.10}
\setlength{\tabcolsep}{2.6pt}
\scriptsize
\begin{tabular}{lccccc}
\toprule
\textbf{benchmark} & \textbf{units} & \textbf{lift} & \textbf{L1} &
\textbf{$\Delta$cos} & \textbf{wins} \\
 & & ours/Meta & ours/Meta & mean$\pm$s & \\
\midrule
Private RAS         & 20 vid. & \best{.265}/.213 & \best{.271}/.300 & $+.0161{\pm}.0025$ & 20/20 \\
GraSP-phase  &  8 cases & \best{.268}/.225 & \best{.268}/.305 & $+.0220{\pm}.0022$ &  8/8 \\
GraSP-step   &  8 cases & \best{.276}/.230 & \best{.268}/.305 & $+.0216{\pm}.0026$ &  8/8 \\
SAR-RARP50   & 20 vid. & \best{.220}/.182 & \best{.268}/.295 & $+.0158{\pm}.0027$ & 20/20 \\
SARAS-ESAD   &  4 vid. & \best{.256}/.202 & \best{.280}/.306 & $+.0157{\pm}.0009$ &  4/4 \\
\bottomrule
\end{tabular}
\end{table}

All downstream results reported above depend, to some extent, on the
task-specific readout. We therefore complement them with a label-free
evaluation based on the masked-prediction objective used during V-JEPA-2.1
pretraining. This experiment asks whether surgical continued pretraining
improves objective-aligned predictive fidelity on surgical video. It is
not a common-space comparison of representation quality, since each
checkpoint is evaluated with its corresponding target encoder and predictor.

For each 16-frame clip at $384^2$ resolution, the 4,608 patch tokens are
randomly partitioned into 1,536 visible context tokens and a complementary
target set. For each checkpoint, the target encoder produces hierarchical
latent representations from four levels of the network; each
1,664-dimensional representation is independently layer-normalized and
concatenated, following the same procedure used during pretraining. The
corresponding online encoder and predictor then estimate these target
representations from the visible context. We report mean cosine similarity,
the L1 distance corresponding to the pretraining objective, and a
shuffled-target cosine baseline obtained by randomly permuting the
correspondence between predicted and target patches. Both checkpoints are
evaluated on the same clips, token partitions, and random seeds, providing
a paired comparison under matched masking conditions.

Raw cosine similarity alone is difficult to compare across checkpoints
because the learned representation spaces exhibit different chance-level
similarities. A representation space in which patch embeddings are
globally more similar can produce a higher shuffled cosine baseline and,
consequently, an inflated matched cosine score. We therefore additionally
report \emph{cosine lift}, defined as the difference between matched cosine
similarity and the checkpoint-specific shuffled baseline. This correction
does not place the two models in a shared latent space, but reduces the
effect of background similarity. The Meta checkpoint exhibits the higher
shuffled baseline, so the separation is larger in cosine lift than in raw
cosine similarity.

Twenty clips per evaluation setting were sampled from OmniRAS-PR, GraSP under
the phase and step sampling protocols, SAR-RARP50, and SARAS-ESAD. We
evaluate the 2B surgically pretrained production checkpoint against the
original Meta 2B checkpoint from which it was initialized. Because clips
sampled from the same source video are not statistically independent,
inferential analyses are conducted on mean differences aggregated at the
independent source-unit level: 20 source videos for our private dataset and SAR-RARP50,
eight surgical cases for GraSP, and four source videos for SARAS-ESAD.

Within this objective-aligned diagnostic, the surgically pretrained checkpoint achieves higher matched cosine similarity, higher cosine lift, and lower L1 prediction error in every evaluation setting. At the independent-unit level, the directional difference favors the surgical checkpoint in all $60/60$ paired comparisons (Fig.~\ref{fig:pretrain-fidelity}). Across benchmarks, the mean per-unit difference in matched cosine similarity ranges from $+0.0157$ to $+0.0220$, with reported paired standard errors of $0.0009$--$0.0026$. The differences remain statistically strong on the benchmarks with 20 independent units; for example, our private dataset yields $t=6.44$ and $p=3.6\times10^{-6}$ under a paired $t$-test. At the individual-clip level, the surgical checkpoint has the higher matched cosine similarity in $99/100$ comparisons, with a single exception among the GraSP-phase clips.

Two qualifications concern statistical interpretation. GraSP-step is not
an independent fifth benchmark because the phase and step settings use
different sampling seeds on the same underlying videos; their similar
results therefore provide a robustness check rather than independent
replication. SARAS-ESAD contains only four independent source videos, so
a sign test cannot attain statistical significance regardless of effect
magnitude. We therefore emphasize the consistent $4/4$ direction rather
than its $p$-value.

Finally, this experiment should not be interpreted as a task-independent
ranking of backbone representations. Because each checkpoint predicts
targets generated by its own target encoder and uses its own predictor, the
evaluation measures checkpoint-specific fidelity to the V-JEPA
masked-prediction objective rather than performance in a shared latent
space. The result instead shows that surgical continued pretraining
consistently improves objective-aligned prediction on held-out surgical
video. Thus, a tie under a frozen downstream probe should not by itself be
taken as evidence that continued pretraining produced no measurable change
in the model's surgical predictive representation.

\section{Discussion}

The adapted surgical encoders obtain the highest score in every task family in this study, and the GraSP results exceed the published end-to-end TAPIS system on both phase and step. The remainder of this discussion decomposes that advantage, since the decomposition attributes less of it to the pretraining campaign than the aggregate ranking would suggest.

The dominant source of improvement is task adaptation rather than the substitution of a surgically continued encoder for a raw one. Last-four-block fine-tuning raises YT-Chole IVT mAP by 4.81--6.56 points, and single-seed ceilings indicate comparable headroom on SAR-RARP50 above the $85.81$--$87.40$ frozen band. It also improves AP$_\text{mean}$ for every V-JEPA variant on SARAS-ESAD, although the frozen OmniRAS 2B remains strongest overall. Frozen probing therefore measures immediate feature accessibility rather than total usable task capacity. Comparisons should report both a frozen probe and a budget-matched partial-adaptation setting.

Surgical \cpt{} is most consistent once the representation is allowed to adapt to workflow tasks. On GraSP, OmniRAS exceeds the corresponding Meta initialization under matched partial fine-tuning for both phase and step recognition. The preferred adaptation regime differs between the two tasks: phase recognition reaches its highest mAP under full fine-tuning, whereas step recognition is strongest with last-four-block fine-tuning. The frozen ordering is reversed on GraSP and scale-dependent on YT-Chole. Source quality may also be consequential, though our one large source effect is not secure: a 2B run including OpenH scored close to seven IVT points below the run excluding it, but its two arms were launcher-mismatched, and the launcher-matched 1B arms are null. At the same time, the 2B \cpt{} model is $6.45$ top-1 accuracy and $7.88$ macro-F1 points below raw on SSv2. Domain adaptation therefore represents a trade-off among surgical transfer, general capability, and compute.

A label-free measurement bears on this decomposition. Replicating the V-JEPA-2.1 masking objective itself on five surgical corpora, with no task head anywhere in the loop, the 2B surgical checkpoint predicts masked surgical targets in its own latent space more accurately than the Meta checkpoint it was initialized from on all 60 paired unit-level comparisons across five evaluation settings (Section~\ref{sec:fidelity}). The frozen probes therefore do not indicate that surgical \cpt{} left the representation unchanged; the label-free result does not, however, identify the readout as the cause of the frozen-probe ties. The corresponding qualification is symmetric. An improved predictive prior is a reason to expect adaptation to begin from a better initialization, but not a demonstration that it terminates at a higher value.

A further consideration is benchmark validity. The custom phase and triplet
tasks probe different temporal granularities, but YT-Chole has only ten batches of ~3 procedur eseach
and our private contribution is an in-house distribution; public GraSP, SAR-RARP50, and SARAS-ESAD
remain necessary external anchors. We use the checkpoint-side pretraining
snapshot as the authoritative manifest and retain case identifiers through
resharding.

The reproducibility of the phase ontology was measured rather than assumed.
Two independent re-annotations of a 10\% sample of YT-Chole agree with
the existing reference at mean $\kappa=0.664$ over all eleven classes at zero
boundary tolerance and $0.807$ at $\pm4$\,s, with the reference sitting inside
rather than outside the rater spread (Section~\ref{sec:irr}). The residual is
structural, in that 82.1\% of all disagreement touches the Calot's-triangle
cluster,
and pooling those five concurrent sub-activities raises the mean to $0.833$,
more than four seconds of tolerance recovers, so raters are selecting between
two valid descriptions of one concurrent activity rather than assigning
different boundaries to the same one. This has two consequences for interpretation. The eleven-class
phase numbers in this paper should be compared against an observed inter-rater agreement of $\kappa$=0.807 at $\pm4$ s. The model-versus-model comparisons are
unaffected, since every encoder is scored against the same reference, but a
model penalized
inside the Calot's cluster is in part being scored on a distinction that
concurrent activity renders intrinsically soft. Second, any surgical phase
ontology that subdivides Calot's triangle subdivides simultaneous activity
rather than a sequence; benchmark designers should either pool those phases or
accept that part of the reported error is attributable to the ontology rather
than to the model.

These results support beginning from a strong general-video checkpoint, reserving budget for partial fine-tuning, screening surgical sources before production training, and reporting immutable manifests together with seeds, checkpoints, head schedules, aggregation, and split policies. For groups planning a continued-pretraining campaign, the three
recommendations derived from the 256-node runs in
Section~\ref{sec:pretraining} state this concretely. They are to verify the
harness before allocating node-hours, evaluate under at least two
readouts, fix bookkeeping conventions before the first launch, and treat small-budget catalog comparisons as exploratory rather than definitive. In our campaign, extending the final configuration to a longer production run was more informative than further branching the small-budget screening runs.




\section{Conclusion}

OmniRAS contributes two annotated robotic-cholecystectomy evaluations, a documented continued-pretraining campaign at up to 256 nodes over a 19-source, majority-robotic surgical catalog of approximately 2,650 hours, and a controlled evaluation of billion-scale predictive video encoders across six surgical downstream tasks. Across 254 seed-level downstream runs, OmniRAS reaches the strongest measured result in every task family considered: YT-Chole action-triplet recognition, OmniRAS-PR phase recognition and temporal segmentation, SAR-RARP50 action recognition, SARAS-ESAD action detection, and GraSP phase and step recognition. These gains extend beyond comparisons with frozen or adapted foundation-model baselines. OmniRAS exceeds the published SAR-RARP50 reference, the best SARAS-ESAD challenge submission, and TAPIS on both GraSP phase and step recognition. Continued surgical pretraining does, however, reduce general-video performance relative to the original 2B initialization: the 36.86M-sample checkpoint scores $53.82$ top-1 on SSv2, $6.45$ points below the raw model.

The experiments also clarify where these gains come from. Within the continued-pretraining recipe, the largest production configuration shows the clearest transfer gains, although its simultaneous catalog change prevents attributing them to compute alone; catalog, sampling, and masking effects remain unresolved at the tested screening budget. Across downstream tasks, frozen comparisons are mixed, whereas partial adaptation yields larger and more consistent gains. GraSP further shows that the preferred adaptation regime depends on the downstream task: phase recognition reaches its highest mAP under full fine-tuning, whereas step recognition is strongest with last-four-block fine-tuning. These results argue against treating either the pretrained encoder or the downstream adaptation protocol as fixed choices when evaluating surgical foundation models.

More broadly, OmniRAS shows that continued pretraining on surgical video can turn a general predictive video model into a strong initialization for surgical tasks without requiring task-specific pretraining for each benchmark. This specialization is not free: it trades some general-video performance for improved surgical transfer, with the magnitude of that trade-off depending on the continued-pretraining configuration. The central result is therefore not a gain on a single dataset, but a consistent transfer advantage across recognition, temporal segmentation, spatial detection, and workflow understanding. We hope this provides both a stronger starting point for robotic-surgery video models and a more rigorous template for evaluating them.
\section{Data and Model Release}
To support reproducibility and further research, we will release all OmniRAS checkpoints together with the two datasets introduced in this work. The two datasets were constructed under two separate IRB-approved protocols, and their release will follow the corresponding institutional, privacy, and data-use requirements.
Access to the models and datasets can be requested through the \href{https://docs.google.com/forms/d/e/1FAIpQLSfXCTF9ISdQ3a_I87wYowRh-0QNLWzCC3ZqQ7wpqbnlgZMN0A/viewform?usp=publish-editor}{\textbf{OmniRAS model and data access request form}}.
The project page is available at \url{https://borgioli.github.io/omniras/}.
\section*{Acknowledgment}

This research used resources of the Argonne Leadership Computing
Facility, a U.S. Department of Energy (DOE) Office of Science user
facility at Argonne National Laboratory, operated under Contract
No. DE-AC02-06CH11357. Argonne National Laboratory's contribution is
based upon work supported by Laboratory Directed Research and
Development (LDRD) funding from Argonne National Laboratory, provided
by the Director, Office of Science, of the U.S. Department of Energy
under Contract No. DE-AC02-06CH11357, through the Convergence
Intelligence Seed Funding Program at the George Crabtree Institute for
Discovery (Project No. 2026-0568).

\appendices

\section{Claude Opus 5 Few shot Triplet Estimation (YT-Chole)}
\begin{table}[h]
\centering
\caption{Comparison between a frontier multimodal model under few-shot prompting
and our task-adapted video representation on the full YT-Chole Triplet
validation set (1,696 clips). IVT mAP is computed over the same 278 supported
triplets for both methods.}
\label{tab:opus5-ytchole}
\setlength{\tabcolsep}{5pt}
\renewcommand{\arraystretch}{0.5}
\begin{tabular}{lcc}
\toprule
Metric &
\textbf{Opus 5} \newline \textbf{few-shot} &
\textbf{OmniRAS} \newline \textbf{FT-full ensemble} \\
\midrule
Tool macro-AP   & 0.722 & \textbf{0.935} \\
Verb macro-AP   & 0.560 & \textbf{0.847} \\
Target macro-AP & 0.351 & \textbf{0.630} \\
IVT mAP         & 0.139 & \textbf{0.426} \\
\bottomrule
\end{tabular}
\end{table}



\ifralappendices

\fi

\end{document}